\documentclass[a4paper,aps,pra,superscriptaddress,showpacs]{revtex4}
\usepackage{graphicx,dcolumn}
\usepackage{amsfonts,amssymb,amsmath}
\usepackage[english]{babel}
\newcommand{\sst}{\scriptscriptstyle}
\begin{document}

\author{Beno{\^\i}t Gr{\'e}maud}
\email{Benoit.Gremaud@spectro.jussieu.fr}
\affiliation{Laboratoire Kastler Brossel, Universit{\'e} Pierre et
Marie Curie, T12, E1 \\
4, place Jussieu, 75252 Paris Cedex 05, France}
\author{Thomas Wellens}
\affiliation{Laboratoire Kastler Brossel, Universit{\'e} Pierre et
Marie Curie, T12, E1 \\
4, place Jussieu, 75252 Paris Cedex 05, France}
\affiliation{Institut Non Lin{\'e}aire de Nice, UMR 6618, 1361 route
des Lucioles, F-06560 Valbonne}
\author{Dominique Delande}
\affiliation{Laboratoire Kastler Brossel, Universit{\'e} Pierre et
Marie Curie, T12, E1 \\
4, place Jussieu, 75252 Paris Cedex 05, France}
\author{Christian Miniatura}
\affiliation{Institut Non Lin{\'e}aire de Nice, UMR 6618, 1361
route des Lucioles, F-06560 Valbonne}

\title{Coherent backscattering in nonlinear atomic media: quantum
  Langevin approach}
\date{\today}

\begin{abstract}
In this theoretical paper, we investigate coherence properties of
the near-resonant light scattered by two atoms exposed to a strong
monochromatic field. To properly incorporate saturation effects,
we use a quantum Langevin approach. In contrast to the standard
optical Bloch equations, this method naturally provides the
inelastic spectrum of the radiated light induced by the quantum
electromagnetic vacuum fluctuations. However, to get the right
spectral properties of the scattered light, it is essential to
correctly describe the statistical properties of these vacuum
fluctuations. Because of the presence of the two atoms, these
statistical properties are not Gaussian : (i) the spatial
two-points correlation function displays a speckle-like behavior
and (ii) the three-points correlation function does not vanish.
We also explain
how to incorporate in a simple way propagation with a
frequency-dependent scattering mean-free path, meaning that the
two atoms are embedded in an average scattering dispersive medium.
Finally we show that saturation-induced nonlinearities strongly
modify the atomic scattering properties and, as a consequence,
provide a source of decoherence in multiple scattering. This is
exemplified by considering the coherent backscattering
configuration where interference effects are blurred by this
decoherence mechanism. This leads to a decrease of the so-called
coherent backscattering enhancement factor.
\end{abstract}

\maketitle
\section{Introduction}

Over the past ten years, cold atomic gases have gradually become a
widely employed and highly tunable tool for testing new ideas in
many areas of quantum physics: quantum phase transitions
(Bose-Einstein condensation, Fermi degenerate gases, Mott-Hubbard
transition)~\cite{bec,fermi,mott-hubbard}, quantum
chaos~\cite{chaos}, applications in metrology~\cite{hsurm},
disordered systems~\cite{cbsat,thierry} to cite a few. In the
latter case, cold atomic vapors act as dilute gases of randomly
distributed atoms multiply scattering an incident monochromatic
laser light. In this case, the scattered light field exhibit a
speckle-like structure due to (multiple) interference between all
possible scattering paths. The key point is that the disorder
average is insufficient to erase all interference effects. This
gives rise to weak or strong localization effects in light
transport depending on the strength of
disorder~\cite{Houches,AkkerMon}. A hallmark of this coherent
transport regime is the coherent backscattering (CBS) phenomenon:
the average intensity multiply scattered off an optically thick
sample is up to twice larger than the average background in a
small angular range around the direction of backscattering,
opposite to the incoming light~\cite{cbs}. This interference
enhancement of the diffuse reflection off the sample is a
manifestation of a two-wave interference. As such, it probes the
coherence properties of the outgoing light~\cite{photon}. The CBS
effect in cold atomic gases has been the subject of extensive
studies in the weak localization regime, both from theoretical and
experimental points of view~\cite{cbsatoms}. In particular,
modifications brought by atoms, as compared to classical
scatterers, for light transport properties (mean-free path,
coherence length, CBS enhancement factor) have been highlighted.
They are essentially due to the quantum internal atomic
structure~\cite{internal,cbsB}.

Another interesting feature of atoms is their ability to display a
nonlinear behavior: the scattered light is no more proportional to
the incident one. This leads to a wide variety of phenomena, like
pattern formation, four-wave mixing, self-focusing effects,
dynamical instabilities,
\emph{etc}~\cite{boyd,prl72GMP,praDHGC,prl85SM}. For a weak
nonlinearity, introducing an intensity-dependent susceptibility is
enough to properly describe these effects, including quantum
properties~\cite{boyd,pra70WGDM,facteur3}, \emph{e.g.} the Kerr
effect (intensity dependence of the refractive index) can be
obtained with a $\chi^{(3)}$ nonlinearity. However, when the
incident intensity is large enough, and this is easily achieved
with atoms, perturbation theories eventually fail and a full
nonlinear treatment is required. For a single two-level atom, the
solution is usually given by the so-called optical Bloch (OB)
equations. Together with the quantum regression theorem, they
allow for a complete description of the spectral properties of the
fluorescence light~\cite{Cohenrouge}. In particular, these
equations show that the atomic nonlinear behavior is intrinsically
linked to the quantum nature of the electromagnetic field. More
specifically, as opposed to classical nonlinear scatterers, the
radiated light exhibits quantum fluctuations characterized by
peculiar time correlation properties. They define a power
spectrum, known as the Mollow triplet, emphasizing inelastic
scattering processes at work in the emission
process~\cite{pr188M,Cohenrouge,ZG}.

However, even if all these aspects are well understood in the case
of a \emph{single} atom exposed to a strong monochromatic
field~\cite{Cohenrouge}, the situation changes dramatically in the
case of a large number of atoms where a detailed analysis
including both quantum nonlinear properties and coherence effects
is still lacking. Until now, the nonlinear coupling between the
atoms and the quantum vacuum fluctuations is either included in a
perturbative scheme~\cite{facteur3,Wellens_long}
or simply described by a classical
noise~\cite{pra46YMC,pra46YMC2,pra51YC,pra52DPGC,pra56B}. In the
dilute regime $\lambda \ll R$ where the light wavelength $\lambda$
is much less than the average particle separation $R$,
one expects the quantum fluctuations to
reduce the degree of coherence of the scattered light.
This will alter not only propagation parameters (mean-free path,
refraction index), but also weak localization corrections to
transport, and the CBS enhancement factor, which is
related to the coherence properties of the scattered
light field~\cite{thierry,photon}. We want here to stress that,
even beyond interference and weak localization phenomena, any
transport property which may be influenced by saturating the
atomic transition deserves a special and necessary study on its
own. The most striking systems falling in this category where both
nonlinear and disordered descriptions are intimately interwoven
are coherent random lasers \cite{cao} where interference
effects lead to localized light modes inside the disordered medium,
comparable to resonator eigenmodes in standard lasers.
Even if, in this case, one
would require an active ({\em i.e.} amplifying) medium, a key
point is the understanding of the mutual effects between multiple
interferences and nonlinear scattering.

In the present paper, we will focus on the rather simple case of
two atoms in vacuum. Our aim is threefold: (i) firstly to properly
calculate quantum correlations between pairs of atoms as a crucial
step towards a better understanding of the physical mechanisms at
work; (ii) secondly to implement a method allowing for a simple
incorporation of frequency-dependent propagation effects; (iii)
finally to understand, in the CBS situation, the modifications
brought by the (quantum) nonlinearity to the interference
properties. We hope that these points, once mastered, can provide
an efficient way to produce realistic computer models to simulate
real experiments. Point (i) alone could easily be solved using the
standard OB method~\cite{pra45VA,prl94SMB}. But the latter almost
becomes useless regarding point (ii), since frequency-dependent
propagation leads to complicated time-correlation functions. From
a numerical point of view, it also leads to such large linear
systems of coupled equations that its practical use is limited up
to only a few atoms, very far from a real experimental situation.
For these reasons, we will rather use the quantum Langevin method
for our purposes. This method not only solves points (i) and (ii),
but also leads to a simple explanation of point (iii), through a
direct evaluation of the quantum noise spectrum. Note however
that, in the absence of any effective medium surrounding the two
atoms, and as long as only the numerical results are concerned
(but not the physical interpretation), the quantum Langevin
approach is completely equivalent to solving the multi-atoms
optical Bloch equations like in~\cite{pra45VA,prl94SMB}.

This paper divides as follows: in section~\ref{oneatom}, the
notations are defined and the quantum Langevin approach is
explained for the single atom case. In section~\ref{twoatoms}, the
method is adapted to the case where two atoms are weakly coupled
by the dipole interaction. The validity and relevance of the
method is controlled by a comparison with a direct calculation
using OB equations. Then, in the CBS configuration, numerical
results for different values of the laser intensity and detuning
are presented and discussed. In particular, possible reasons for
the reduction of the enhancement factor are put forward.

\section{Single two-level atom case}
\label{oneatom}
\subsection{Time-domain approach}

We consider an atom with a zero angular momentum electronic ground
state ($J_g=0$) exposed to a monochromatic light field. The light
field frequency $\omega_L$ is near-resonant with an optical dipole
transition connecting this ground state to an excited state with
angular momentum $J_e=1$. The angular frequency separation between
these two states is $\omega_0$ and the natural linewidth of the
excited state is $\Gamma$. We will denote hereafter by $\delta =
\omega_L-\omega_0$ the laser detuning. The ground state is denoted
by $|0\,0\rangle$ while the excited states are denoted by
$|1\,m_e\rangle$, with $m_e=-1,0,1$ the Zeeman magnetic quantum
number. As we assume no magnetic field to be present throughout this
paper, the excited state is triply degenerate.

In the Heisenberg picture, this two-level atom is entirely
characterized by the following set of 16 time-dependent
operators:

\begin{equation}
\Pi^g=|0\,0\rangle\langle 0\,0| \quad ; \quad
\Pi^e_{m_e\,m'_e}=|1\,m_e\rangle\langle 1\,m'_e| \quad ; \quad
\mathcal{D}^+_{m_e}=|1\,m_e\rangle\langle 0\,0| \quad ; \quad
\mathcal{D}^-_{m_e}=|0\,0\rangle\langle 1\,m_e|
\end{equation}
The atomic operators obey the completeness constraint
\begin{equation}
\label{constraint}
\openone=\Pi^g + \Pi^e
\end{equation}
where $\Pi^g$ and $\Pi^e = \sum_{m_e} \Pi^e_{m_e\,m_e}$ are the
ground and excited state atomic population operators.

The full atom-field Hamiltonian $\mathcal{H}$ is the sum of the
free atom Hamiltonian $\mathcal{H}_A = \hbar\omega_0 \Pi^e$, of
the free quantized field Hamiltonian $\mathcal{H}_F =
\sum_{\textbf{k},\boldsymbol{\epsilon} \perp \textbf{k}}
\hbar\omega_\textbf{k} \, a^\dag_{\textbf{k}\boldsymbol{\epsilon}}
a_{\textbf{k}\boldsymbol{\epsilon}}$ and of the dipolar
interaction $\mathcal{V} = - \mathbf{d} \cdot (\mathbf{E}_L +
\mathbf{E}_V)$ between the atomic dipole $\mathbf{d}$, the
classical laser field $\mathbf{E}_L$ and the quantum
electromagnetic vacuum field $\mathbf{E}_V$. Performing the usual
approximations of quantum optics, \emph{i.e.} neglecting
non-resonant terms (rotating wave approximation) and assuming
Markov-type correlations between the atomic operators and the
vacuum field, one obtains the quantum Langevin equations
controlling the time evolution of any atomic observable
$\mathcal{O}$ in the rotating frame~\cite{pra46YMC,Cohenrouge}:
\begin{equation}
\label{langevin1}
\frac{d\mathcal{O}}{dt}=i\delta_L[\mathcal{O},\Pi^e]
-\frac{i}{2}\sum_q(-1)^q[\mathcal{O},\mathcal{D}^+_q]\Omega^{L+}_{-q}(\textbf{R})
-\frac{i}{2}\sum_q[\mathcal{O},\mathcal{D}^-_q]\Omega^{L-}_{q}(\textbf{R})
-\frac{\Gamma}{2}\left(\mathcal{O}\Pi^e+\Pi^e\mathcal{O}\right)
+\Gamma\sum_q\mathcal{D}^+_q\mathcal{O}\mathcal{D}^-_q+
\mathcal{F}_{\mathcal{O}}(\textbf{R},t),
\end{equation}
where $\Omega^{L+}_q$ (resp. $\Omega^{L-}_q$) are the components
of the Rabi frequency of the positive (resp. negative) frequency
parts of the incident laser beam, i.e.
$\hbar\mathbf{\Omega}=-d\mathbf{E}$ where $d$ is the
dipole strength.
Finally $\mathcal{F}_{\mathcal{O}}(t)$ is the Langevin force
depicting the effects of the quantum fluctuations of the vacuum
electromagnetic field and reads as follows:
\begin{equation}
\label{langevinforce}
\mathcal{F}_{\mathcal{O}}(t)=-\frac{i}{2}\sum_q(-1)^q
[\mathcal{O},\mathcal{D}^+_q] \Omega^{0+}_{-q}(\textbf{R},t)
-\frac{i}{2}\sum_q\Omega^{0-}_{q}(\textbf{R},t)[\mathcal{O},\mathcal{D}^-_q],
\end{equation}
where $\Omega^{0+}(\textbf{R},t)$ is the vacuum Rabi field
operator
\begin{equation}
\mathbf{\Omega}^{0+}(\textbf{R},t)=-\frac{2id}{\hbar}\sum_{\textbf{k},
\boldsymbol{\epsilon}\perp\textbf{k}}\mathcal{E}(\omega)
\boldsymbol{\epsilon}\,a_{\textbf{k}\boldsymbol{\epsilon}}(t_0)
e^{i\textbf{k}\cdot\textbf{R}-i(\omega-\omega_L)(t-t_0)}
\end{equation}
with $t_0$ an initial time far in the past. From the preceding
expression, one can calculate the time correlation functions of
the vacuum field~\cite{Cohengris}:
\begin{equation}
(-1)^q[\Omega^{0+}_{-q}(\textbf{R},t),\Omega^{0-}_{q'}(\textbf{R},t')]=
4\Gamma\delta_{q\,q'}f(t-t'),
\end{equation}
where $f(\tau)$ in a function centered around $\tau=0$, whose
width $\tau_c$ is much smaller than any characteristic atomic
timescale (i.e. $\tau_c\ll \omega_0^{-1}\ll\Gamma^{-1}$) and whose
time integral is equal to unity. Thus, hereafter, $f(\tau)$ will
be safely replaced by a $\delta$-function: $f(\tau) \to
\delta(\tau)$.

The time evolution for the expectation values is obtained by
averaging over the initial density matrix $\sigma(t_0)$, i.e.,
$\langle\mathcal{O}(t)\rangle=\mathrm{Tr}(\mathcal{O}(t)\sigma(t_0))$.
Since the atom and the vacuum field are supposed to be decoupled
initially, $\sigma(t_0)$ is simply
$\sigma_{at}(t_0)\otimes|0\rangle\langle 0|$ ($|0\rangle$ being
the vacuum field state). Because of the normal ordering, one
immediately gets:
\begin{equation}
\langle\mathcal{F}_{\mathcal{O}}(t)\rangle=0,
\end{equation}
and the time correlation functions of the Langevin forces:
\begin{equation}
\langle\mathcal{F}_{\mathcal{O}}(t)\mathcal{F}_{\mathcal{O}'}(t')\rangle=
-\Gamma\left\langle\sum_q [\mathcal{O}(t),\mathcal{D}^+_q(t)]
[\mathcal{O}'(t'),\mathcal{D}^-_q(t')]\right\rangle \delta(t-t').
\label{corr_Langevin}
\end{equation}

The physical picture of the quantum Langevin approach is to
represent quantum fluctuations by a fluctuating force acting on
the system, in analogy with the usual Brownian motion. Not
surprisingly, this leads to a diffusive-like behavior of
expectation values. More precisely, because of the
$\delta$-function in Eq.~(\ref{corr_Langevin}), we can set $t'=t$
for the atomic operators and we finally obtain in the stationary
regime $t \gg t_0$:
\begin{equation}
\langle\mathcal{F}_{\mathcal{O}}(t)\mathcal{F}_{\mathcal{O}'}(t')\rangle=
\frac{\Gamma}{4} \, D_{\mathcal{O}\,\mathcal{O}'}\, \delta(t-t'),
\end{equation}
where $D$ is a matrix of diffusion constants depending only on the stationary
values of the atomics operators. The stationary hypothesis also
results from the fact that these correlation functions only depend
on the time difference $t-t'$.

From this, it is possible to prove that the quantum regression
theorem applies~\cite{CR92,Cohenrouge}, allowing for the
calculation of two-times correlation functions of the atomic
operators and of their expectation values. From their Fourier
transforms, one can obtain the spectrum of the radiated light.
But, for the reasons mentioned in the introduction, we will
explain how these properties can be obtained in a much simpler way
by directly translating the Langevin equations in the Fourier
domain~\cite{CR92}.

\subsection{Frequency-domain approach}

First, because of the constraint~\eqref{constraint}, only 15
atomic operators are actually independent. More specifically, we
will use the following set, denoted by the column vector
$\mathbf{X}$:
\begin{equation}
\textbf{X}\left\{
\begin{aligned}
\Pi^z_{m_e}&=\frac{1}{2}\left[\Pi^e_{m_e\,m_e}-\Pi^g\right]\\
\Pi^e_{m_e\,m'_e}&=|1\,m_e\rangle\langle 1\,m'_e|\qquad m_e\neq m'_e\\
\mathcal{D}^+_{m_e}&=|1\,m_e\rangle\langle 0\,0|\\
\mathcal{D}^-_{m_e}&=|0\,0\rangle\langle 1\,m_e|
\end{aligned}\right..
\label{def_X}
\end{equation}
The Langevin equations for $\mathbf{X}$ then formally read as follows:
\begin{equation}
\label{Langevin2}
\frac{d}{dt}\mathbf{X}(t)=M\mathbf{X}(t)+\mathbf{L}+\mathbf{F}(t),
\end{equation}
where $M$ is a time-independent matrix depending on the laser Rabi
frequency $\Omega^{L\pm}$, $\mathbf{L}$ is a constant vector
scaling with $\Gamma$ and $\mathbf{F}(t)$  is a vector
characterizing the Langevin forces at work on the atom (for
simplicity, we have dropped the explicit position dependence). The
stationary expectation values are then simply given by:
\begin{equation}
\label{stationary}
\langle \mathbf{X}\rangle=-M^{-1}\mathbf{L}.
\end{equation}

Using Kubo's notations, the Fourier transforms of the different
quantities are defined as follows:
\begin{equation}
\begin{aligned}
f[\Delta]&=\int dt f(t)e^{i\Delta t}\\
f(t)&=\int \frac{d\Delta}{2\pi} f[\Delta]e^{-i\Delta t},
\end{aligned}
\end{equation}
leading to the Langevin equations in the frequency domain:
\begin{equation}
\label{Langevinfreq}
\left(-i\Delta\openone-M\right)\mathbf{X}[\Delta]=2\pi\delta[\Delta]\mathbf{L}+
\mathbf{F}[\Delta].
\end{equation}

Introducing the Green's function
$G[\Delta]=\left(-i\Delta\openone-M\right)^{-1}$, the solution of
the preceding equations simply reads:
\begin{equation}
\label{sollanfre}
\mathbf{X}[\Delta]=G[\Delta]\left(2\pi\delta[\Delta]\mathbf{L}+
\mathbf{F}[\Delta]\right).
\end{equation}

Using $G[0]=-M^{-1}$ and (\ref{stationary}), this solution
separates into a non-fluctuating part $\mathbf{X}_L[\Delta]$ and a
fluctuating (frequency-dependent) part $\mathbf{X}_F[\Delta]$:
\begin{equation}
\left\{
\begin{aligned}
\mathbf{X}_L[\Delta]&=2\pi\delta[\Delta]\langle\mathbf{X}\rangle\\
\mathbf{X}_F[\Delta]&=G[\Delta]\mathbf{F}[\Delta]
\end{aligned}\right..
\end{equation}

From the linearity of the Fourier transform, we still have
$\langle\mathbf{F}[\Delta]\rangle=\mathbf{0}$ implying
$\langle\mathbf{X}_F[\Delta]\rangle=\mathbf{0}$. The time
correlation functions for the Langevin force components, Eq.~(\ref{corr_Langevin}), become:
\begin{equation}
\label{corr11} \langle
\textbf{F}_p[\Delta']\textbf{F}_q[\Delta]\rangle=2\pi\delta[\Delta'+\Delta]D_{pq}.
\end{equation}
where the $2\pi\delta[\Delta'+\Delta]$ function is a direct
consequence of the time-translation invariance, \emph{i.e.} that
we calculate the correlation functions in the stationary regime.
This implies that the correlation function for the components of
$\mathbf{X}_F$ in the frequency domain are:
\begin{equation}
\langle \big(\mathbf{X}_F[\Delta']\big)_p \,
\big(\mathbf{X}_F[\Delta]\big)_q \rangle = 2\pi
\delta[\Delta+\Delta'] \, \big(G \, D \, ^t\!G \big)_{pq}
\end{equation}
where the superscript $t$ means matrix transposition.

The field radiated at frequency $\Delta$ by the atom at a distance
$r\gg\lambda$ (far-field regime) reads as follows:
\begin{equation}
\label{propvac}
\Omega^+_q[\Delta]=-\frac{3}{2}\Gamma\,\mathcal{P}^{\textbf{r}}_{qq'}
\,\mathcal{D}^-_{q'}[\Delta]\frac{e^{ikr}}{kr},
\end{equation}
where we use implicit sum over repeated indices and where
$\mathcal{P}^{\textbf{r}}$ is the projector onto the plane
perpendicular to vector $\textbf{r}$:
\begin{equation}
\mathcal{P}^{\textbf{r}}_{qq'}=
\bar{\boldsymbol{\epsilon}}_{q}\mathcal{P}^{\textbf{r}}\boldsymbol{\epsilon}_{q'}
=\bar{\boldsymbol{\epsilon}}_{q}
\left(\openone-\frac{\textbf{r}{}^{\;\;t}\!\textbf{r}}{r^2}\right)
\boldsymbol{\epsilon}_{q'}=\delta_{qq'}-(-1)^q\frac{\textbf{r}_{-q}
\textbf{r}_{q'}}{r^2},
\end{equation}
where the bar denotes complex conjugation and where
$(\textbf{r}{}^{\;\;t}\!\textbf{r})$ is a dyadic tensor.

The correlation functions $\langle \Omega^{-}_{q'}[\Delta']
\Omega^{+}_q[\Delta]\rangle$  of the light emitted by the atoms is then
proportional to $\langle \mathcal{D}^{+}_{q'}[\Delta']
\mathcal{D}^{-}_q[\Delta]\rangle$ and read as follow:
\begin{equation}
\langle \Omega^{-}_{q'}[\Delta']\Omega^{+}_q[\Delta]\rangle\propto
(2\pi)^2\delta[\Delta]\delta[\Delta']\langle
\mathcal{D}^{+}_{q'}\rangle\langle\mathcal{D}^{-}_q\rangle
+2\pi\delta[\Delta'+\Delta]\sum_{p'p}G_{i'p'}(\Delta')G_{ip}(\Delta)D_{p'p},
\end{equation}
where the index $i$ (resp. $i'$) corresponds to
$\mathcal{D}^{-}_q$ (resp. $\mathcal{D}^{+}_{q'}$). The
non-fluctuating part gives rise to a spectral component of the
emitted light at exactly the incident laser frequency and is thus
naturally called the \textit{elastic} part. The fluctuating part
gives rise to the \textit{inelastic} Mollow triplet
spectrum~\cite{pra5M}, whose properties (position and width of the
peaks) are given by the poles of $G[\Delta]$, \emph{i.e.} by the
complex eigenvalues of $M$. Actually, we simply recover the
results of the quantum regression theorem, which states that the
atomic time correlation functions evolve with the same equations
than the expectation values
$\dot{\langle\mathbf{X}\rangle}=M\langle\mathbf{X}\rangle+\mathbf{L}$~\cite{Cohenrouge,pr188M}.

\section{Two-atom case}
\label{twoatoms}
\subsection{Optical Bloch equations}
\label{equivalence_ob}
We now consider two isolated atoms, located at fixed positions
$\textbf{R}_1$ and $\textbf{R}_2$. Defining
$\textbf{R}=\textbf{R}_2-\textbf{R}_1 = R \, \textbf{u}$ (with
$R=|\textbf{R}|$ and $\textbf{u}$ the unit vector joining atom 1
to atom 2), we assume the far-field condition $R\gg\lambda$ to
hold. We also assume that $R$ is sufficiently small for the light
propagation time $R/c$ to be much smaller than any typical atomic
timescales $(\Gamma^{-1},\delta^{-1},\Omega_L^{-1}$). In this
regime, all quantities involving the two atoms
are to be computed at the same time $t.$
The contribution of the atom-atom dipole interaction in the
Langevin equation for any
atomic operator $\mathcal{O}$ reads:
\begin{equation}
\label{langevin12}
{\left.\frac{d\mathcal{O}}{dt}\right|}_{\text{dip.}}=i\frac{3\Gamma}{4}
\left\{\left(\left[\mathcal{O},\mathcal{D}^{1+}_q\right]
\mathcal{P}^{\textbf{R}}_{qq'}\mathcal{D}^{2-}_{q'}+
\left[\mathcal{O},\mathcal{D}^{2+}_q\right]\mathcal{P}^{\textbf{R}}_{qq'}
\mathcal{D}^{1-}_{q'}\right)\frac{e^{ikR}}{kR}
+\left(\mathcal{D}^{1+}_q\mathcal{P}^{\textbf{R}}_{qq'}
\left[\mathcal{O},\mathcal{D}^{2-}_{q'}\right]+
\mathcal{D}^{2+}_q\mathcal{P}^{\textbf{R}}_{qq'}
\left[\mathcal{O},\mathcal{D}^{1-}_{q'}\right]\right)
\frac{e^{-ikR}}{kR}\right\}.
\end{equation}

In the OB equations, the two-atom system is entirely described by
the set of 256 operators $X_{ij}$ made of all possible products
$X_i^{1}X_j^{2}$. The stationary expectation values $\langle
X_{ij}\rangle$ are then obtained as solutions of a linear system
resembling equation~\eqref{stationary}. This is the
approach used in~\cite{prl94SMB}, where such optical Bloch equations
are solved.

Since the two atoms are far enough from each other, the
electromagnetic field radiated by one atom onto the other can be
treated as a perturbation with respect to the incident laser
field. More precisely, the  solutions $\langle X_{ij}\rangle$ can
be expanded up to second order in powers of $g$ and $\bar{g}$:
\begin{equation}
\langle X_{ij}\rangle=\langle X_{ij}\rangle^{(0)}+ g\,\langle
X_{ij}\rangle^{(g)}+\bar{g}\,\langle X_{ij}\rangle^{(\bar{g})}+
g\bar{g}\,\langle X_{ij}\rangle^{(g\bar{g})}+ g^2 \,\langle
X_{ij}\rangle^{(gg)} + \bar{g}^2 \,\langle
X_{ij}\rangle^{(\bar{g}\bar{g})}
\end{equation}
where the complex coupling constant $g$ is:
\begin{equation}
g=i\frac{3\Gamma}{2}\frac{\exp{(ikR)}}{kR}
\end{equation}

In fact, it will be shown below that both terms in $g^2$ and
$\bar{g}^2$ give a vanishing contribution to the coherent
backscattering signal.

As explained in the introduction, this approach has two drawbacks:
(i) the solutions obtained in this way are global and, thus, do
not provide a simple understanding of the properties of the
emitted light; (ii) when the two atoms are embedded in a medium
whose susceptibility strongly depends on the frequency, the field
radiated by one atom onto the other at a given time $t$ now
depends on the atomic operators of the first atom at earlier times
(since retardation effects become frequency dependent). Time
correlation functions in the dipole interaction then  explicitly
show up.

\subsection{Langevin approach}

The Langevin equations for the two sets of atomic operators
$\mathbf{X}^{\alpha}$, with $\alpha=1,2$, read formally:
\begin{equation}
\dot{\mathbf{X}}^{\alpha}=M^{\alpha}\mathbf{X}^{\alpha}+\mathbf{L}+
\mathbf{F}^{\alpha}+
gT^{q+}\mathbf{X}^{\alpha}\mathcal{P}^{\textbf{R}}_{qq'}
\mathcal{D}^{\beta-}_{q'}+\bar{g}\mathcal{D}^{\beta+}_q
\mathcal{P}^{\textbf{R}}_{qq'}T^{q'-}\mathbf{X}^{\alpha},
\end{equation}
where $\beta$ denotes the other atom and where $T^{q\pm}$ are
$15\times15$ matrices defined by
$\left[X_i,\mathcal{D}^{\pm}_q\right]=\pm 2T^{q\pm}_{ij}X_j$.
Taking the Fourier transform of these equations, one gets:
\begin{equation}
\label{langevinalpha}
\mathbf{X}^{\alpha}[\Delta]=G^{\alpha}[\Delta]
\left(2\pi\delta[\Delta]\mathbf{L}+\mathbf{F}^{\alpha}[\Delta]\right)
+gG^{\alpha}[\Delta]T^{q+}\mathcal{P}^{\textbf{R}}_{qq'}
\left(\mathbf{X}^{\alpha}\otimes\mathcal{D}^{\beta-}_{q'}\right)[\Delta]
-\bar{g}G^{\alpha}[\Delta]\mathcal{P}^{\textbf{R}}_{qq'}T^{q'-}
\left(\mathcal{D}^{\beta+}_q\otimes\mathbf{X}^{\alpha}\right)[\Delta],
\end{equation}
where $\otimes$ is the convolution operator:
\begin{equation}
\left(A\otimes B\right)[\Delta]=\frac{1}{2\pi}\iint d\Delta_1d\Delta_2
\delta[\Delta_1+\Delta_2-\Delta]A[\Delta_1]B[\Delta_2].
\end{equation}
Introducing, for simplicity, the following notations:
\begin{equation}
\left\{\begin{aligned}
\mathbf{X}^{\alpha^{(0)}}[\Delta]&=G^{\alpha}[\Delta]
\left(2\pi\delta[\Delta]\mathbf{L}+\mathbf{F}^{\alpha}[\Delta]\right)\\
G^{\alpha^+_q}[\Delta]&=G^{\alpha}[\Delta]T^{q'+}
\mathcal{P}^{\textbf{R}}_{q'q}\\
G^{\alpha^-_q}[\Delta]&=G^{\alpha}[\Delta]T^{q'-}
\mathcal{P}^{\textbf{R}}_{qq'}
\end{aligned}\right.,
\end{equation}
equation~\eqref{langevinalpha} becomes:
\begin{equation}
\label{langevin2}
\mathbf{X}^{\alpha}[\Delta]=\mathbf{X}^{\alpha^{(0)}}[\Delta]+
gG^{\alpha_q^+}[\Delta]\left(\mathbf{X}^{\alpha}\otimes
\mathcal{D}^{\beta-}_{q}\right)[\Delta]-
\bar{g}G^{\alpha_q^-}[\Delta]\left(\mathcal{D}^{\beta+}_q\otimes
\mathbf{X}^{\alpha}\right)[\Delta],
\end{equation}
from which one gets the expansion in power of $g$ and $\bar{g}$ (up to
$g\bar{g}$) for
the atomic operators:
\begin{equation}
\label{opexp}
\begin{aligned}
X_i^{\alpha}[\Delta]&=X_i^{\alpha^{(0)}}[\Delta]
+gG_{ij}^{\alpha^+_q}[\Delta] \bigl(X_j^{\alpha^{(0)}}\otimes
\mathcal{D}^{{\beta-}^{(0)}}_q\bigr)[\Delta]
-\bar{g}G_{ij}^{\alpha^-_q}[\Delta]
\bigl(\mathcal{D}^{{\beta+}^{(0)}}_q\otimes
X_j^{\alpha^{(0)}}\bigr)[\Delta]\\
&-g\bar{g}\biggl\{G_{ij}^{\alpha^+_q}[\Delta]\left(X_j^{\alpha^{(0)}}
\otimes G^{\beta^-_p}_{\mathcal{D}_q^-j'}
\left(\mathcal{D}^{{\alpha+}^{(0)}}_p \otimes
X^{\beta^{(0)}}_{j'}\right)\right)[\Delta]
+G_{ij}^{\alpha^+_q}[\Delta]\left(G^{\alpha^-_p}_{jj'}
\left(\mathcal{D}^{{\beta+}^{(0)}}_p\otimes
X_{j'}^{\alpha^{(0)}}\right)\otimes
\mathcal{D}^{{\beta-}^{(0)}}_q\right)[\Delta]
\biggr.\\
&\phantom{+g\bar{g}}\biggl.\quad+G_{ij}^{\alpha^-_q}[\Delta]
\left(\mathcal{D}^{{\beta+}^{(0)}}_q\otimes
G^{\alpha^+_p}_{jj'}\left(X_{j'}^{\alpha^{(0)}}\otimes
\mathcal{D}^{{\beta-}^{(0)}}_p\right)\right)[\Delta]
+G_{ij}^{\alpha^-_q}[\Delta]\left(G^{\beta^+_p}_{\mathcal{D}_q^-j'}
\left(X_{j'}^{\beta^{(0)}}\otimes
\mathcal{D}^{{\alpha-}^{(0)}}_p\right)\otimes
X_j^{\alpha^{(0)}}\right)[\Delta] \biggr\}.
\end{aligned}
\end{equation}
Two-body terms expansions, obtained from Eq.~\eqref{opexp}, read
as follows:
\begin{equation}
\begin{aligned}
\label{corrggb} {X}_{i'}^{\beta}[\Delta']{X}_i^{\alpha}[\Delta]&=
{X}_{i'}^{\beta^{(0)}}[\Delta']{X}_i^{\alpha^{(0)}}[\Delta]\\
&+g\biggl\{{X}_{i'}^{\beta^{(0)}}[\Delta']
G_{ij}^{\alpha^+_q}[\Delta] \bigl(X_j^{\alpha^{(0)}}\otimes
\mathcal{D}^{{\beta-}^{(0)}}_q\bigr)[\Delta]
+G_{i'j'}^{\beta^+_q}[\Delta'] \bigl(X_{j'}^{\beta^{(0)}}\otimes
\mathcal{D}^{{\alpha-}^{(0)}}_q\bigr)[\Delta']
{X}_i^{\alpha^{(0)}}[\Delta]\biggr\}\\
&-\bar{g}\biggl\{{X}_{i'}^{\beta^{(0)}}[\Delta']
G_{ij}^{\alpha^-_q}[\Delta]
\bigl(\mathcal{D}^{{\beta+}^{(0)}}_q\otimes
X_j^{\alpha^{(0)}}\bigr)[\Delta]+ G_{i'j'}^{\beta^-_q}[\Delta']
\bigl(\mathcal{D}^{{\alpha+}^{(0)}}_q\otimes
X_{j'}^{\beta^{(0)}}\bigr)[\Delta']
{X}_i^{\alpha^{(0)}}[\Delta]\biggr\}\\
&-g\bar{g}\biggl\{\text{see appendix~\ref{ggbar}}\biggr\}\\
{X}_{i'}^{\alpha}[\Delta']{X}_i^{\alpha}[\Delta]&=
{X}_{i'}^{\alpha^{(0)}}[\Delta']{X}_i^{\alpha^{(0)}}[\Delta]\\
&+g\biggl\{{X}_{i'}^{\alpha^{(0)}}[\Delta']
G_{ij}^{\alpha^+_q}[\Delta] \bigl(X_j^{\alpha^{(0)}}\otimes
\mathcal{D}^{{\beta-}^{(0)}}_q\bigr)[\Delta]
+G_{i'j'}^{\alpha^+_q}[\Delta'] \bigl(X_{j'}^{\alpha^{(0)}}\otimes
\mathcal{D}^{{\beta-}^{(0)}}_q\bigr)[\Delta']
{X}_i^{\alpha^{(0)}}[\Delta]\biggr\}\\
&-\bar{g}\biggl\{{X}_{i'}^{\alpha^{(0)}}[\Delta']
G_{ij}^{\alpha^-_q}[\Delta]
\bigl(\mathcal{D}^{{\beta+}^{(0)}}_q\otimes
X_j^{\alpha^{(0)}}\bigr)[\Delta] +G_{i'j'}^{\alpha^-_q}[\Delta']
\bigl(\mathcal{D}^{{\beta+}^{(0)}}_q\otimes
X_{j'}^{\alpha^{(0)}}\bigr)[\Delta']
{X}_i^{\alpha^{(0)}}[\Delta]\biggr\}\\
&-g\bar{g}\biggl\{\text{see appendix~\ref{ggbar}}\biggr\}.
\end{aligned}
\end{equation}
Obviously, the power expansion of the expectation values can be derived
from the quantum average of the preceding equations, but not as easily
as it seems. Indeed, if one formally writes:
\begin{equation}
\left\{\begin{aligned}
X_{i'}^{\alpha'}[\Delta']{X}_i^{\alpha}[\Delta]&=\sum_{ab}O(a,b) g^a\bar{g}^b\\
\langle X_{i'}^{\alpha'}[\Delta']{X}_i^{\alpha}[\Delta]\rangle
&=\sum_{ab}C(a,b) g^a\bar{g}^b
\end{aligned}\right.,
\end{equation}
then $C(a,b)$ is not simply equal to
$\langle O(a,b)\rangle$. Actually, $C(a,b)$ depends on all
$\langle O(a',b')\rangle$ for $(a',b')\le(a,b)$, and this for two
reasons:
\begin{itemize}
\item[$\bullet$] for a given atom $\alpha$, the frequency correlation
  functions $\langle F_p^{\alpha}[\Delta']F_q^{\alpha}[\Delta]\rangle$
  are given by $2\pi\delta[\Delta'+\Delta]D_{pq}$, where $D_{pq}$
  depends on the stationary values. But the latter are modified by the
  second atom and, thus, must also be expanded in power of $g$ and
  $\bar{g}$. This implies, for example, that the first term
  ${X}_{i'}^{\alpha^{(0)}}[\Delta']{X}_i^{\alpha^{(0)}}[\Delta]$ in
the expansion of ${X}_{i'}^{\alpha}[\Delta']{X}_i^{\alpha}[\Delta]$
 (Eq.~\eqref{corrggb}) will contribute to all coefficients of
$\langle {X}_{i'}^{\alpha}[\Delta']{X}_i^{\alpha}[\Delta]\rangle$.

\item[$\bullet$] the Langevin forces acting on two different atoms are
  correlated since they both originate from the vacuum quantum
  field. More precisely, their frequency correlation
  functions depend on their relative distance. This dependence
  is analogous to the correlation function of a speckle pattern
  (resulting from the random superposition of plane waves with
  the same wavelength but arbitrary directions):
\begin{equation}
\begin{aligned}
\label{corrfafab} \langle
F_{i'}^{\beta}[\Delta']F_i^{\alpha}[\Delta]\rangle&=
2\pi\delta[\Delta'+\Delta]\frac{3}{2}\Gamma\frac{\sin{kR}}{kR}
T_{i'j'}^{q'+}\mathcal{P}^{\textbf{R}}_{q'q}T_{ij}^{q-}
\langle X_{j'}^{\beta}X_j^{\alpha}\rangle\\
&=-\frac{1}{2}\biggl(g+\bar{g}\biggr)2\pi\delta[\Delta'+\Delta]
T_{i'j'}^{q'+}\mathcal{P}^{\textbf{R}}_{q'q}T_{ij}^{q-}
\langle X_{j'}^{\beta}X_j^{\alpha}\rangle\\
&=-\frac{1}{2}\biggl(g+\bar{g}\biggr)2\pi\delta[\Delta'+\Delta]
D^{\beta\alpha}_{i'i}.
\end{aligned}
\end{equation}
Thus, terms like $ {X}_{i'}^{\beta^{(0)}}[\Delta']
\bigl(X_j^{\alpha^{(0)}}\otimes
\mathcal{D}^{{\beta-}^{(0)}}_q\bigr)[\Delta]$ appearing in
equation~\eqref{corrggb} will also contribute to higher-order
coefficients in the power expansion of
$\langle {X}_{i'}^{\beta}[\Delta']{X}_i^{\alpha}[\Delta]\rangle$. One must note
that, when $R\rightarrow 0$,
$\mathcal{P}^{\textbf{R}}_{q'q}\rightarrow\frac{2}{3}\delta_{q'q}$
and one recovers the single atom correlation functions given by
Eq.~\eqref{corr11}, which emphasizes the consistency of the
present approach.
\end{itemize}

Despite these subtleties, it is nevertheless possible to calculate
power expansions of the atomic correlation functions. More
precisely, in order to emphasize the validity of the present
approach, we will compare the results obtain from the OB equations
and from the Langevin approach. Indeed from the atomic correlation
functions, the stationary solutions can be calculated by inverse
Fourier transform as follows:
\begin{equation}
\langle X_{i'}^{\alpha}X_{i}^{\alpha'}\rangle=\frac{1}{(2\pi)^2}\iint
d\Delta'd\Delta\langle X_{i'}^{\alpha}[\Delta']X_{i}^{\alpha'}[\Delta]\rangle.
\end{equation}

As a specific example, the coefficient proportional to $g$ in the
perturbative expansion  of
$\langle{X}_{i'}^{\beta}[\Delta']{X}_i^{\alpha}[\Delta]\rangle$ is
given by:
\begin{equation}
\begin{aligned}
\langle{X}_{i'}^{\beta}[\Delta']{X}_i^{\alpha}[\Delta]\rangle^{(g)}&=
\underline{\langle{X}_{i'}^{\beta^{(0)}}[\Delta']
{X}_i^{\alpha^{(0)}}[\Delta]\rangle^{(g)}}
+\langle{X}_{i'}^{\beta^{(0)}}[\Delta']
G_{ij}^{\alpha^+_q}[\Delta] \bigl(X_j^{\alpha^{(0)}}\otimes
\mathcal{D}^{{\beta-}^{(0)}}_q\bigr)[\Delta]\rangle^{(0)}\\
&+\langle G_{i'j'}^{\beta^+_q}[\Delta']
\bigl(X_{j'}^{\beta^{(0)}}\otimes
\mathcal{D}^{{\alpha-}^{(0)}}_q\bigr)[\Delta']
{X}_i^{\alpha^{(0)}}[\Delta]\rangle^{(0)}\\
&=\underline{G^{\beta}_{i'j'}[\Delta']G^{\alpha}_{ij}[\Delta]
\langle
F_{j'}^{\beta}[\Delta']F_{j}^{\alpha}[\Delta]\rangle^{(g)}}+
G_{ij}^{\alpha^+_q}[\Delta]\langle X_j^{\alpha^{(0)}}\rangle
\langle{X}_{i'}^{\beta^{(0)}}[\Delta']
\mathcal{D}^{{\beta-}^{(0)}}_q[\Delta]\rangle^{(0)}\\
&+G_{i'j'}^{\beta^+_q}[\Delta'] \langle
X_{j'}^{\beta^{(0)}}\rangle \langle
\mathcal{D}^{{\alpha-}^{(0)}}_q[\Delta']
{X}_i^{\alpha^{(0)}}[\Delta]\rangle^{(0)},
\end{aligned}
\end{equation}
where we have used the fact that terms like $\langle
X^{\alpha^{(0)}}X^{\beta^{(0)}}\rangle^{(0)}$ (\emph{i.e.} zeroth
order) actually factorize into $\langle X^{\alpha}\rangle \langle
X^{\beta}\rangle$ since
  their fluctuating parts necessarily give rise to higher orders in $g$
  and $\bar{g}$, see Eq.~\eqref{corrfafab}. The underlined terms
  correspond to the non-vanishing correlations of the quantum vacuum
  fluctuations evaluated at the two atom positions.

Finally, separating elastic and inelastic part, one gets:
\begin{equation}
\begin{aligned}
\langle{X}_{i'}^{\beta}[\Delta']{X}_i^{\alpha}[\Delta]\rangle^{(g)}&=
(2\pi)^2\delta[\Delta']\delta[\Delta]
\biggl(G_{ij}^{\alpha^+_q}[0]\langle X_j^{\alpha^{(0)}}\rangle
\langle{X}_{i'}^{\beta^{(0)}}\rangle
\langle\mathcal{D}^{{\beta-}^{(0)}}_q\rangle
+G_{i'j'}^{\beta^+_q}[0] \langle X_{j'}^{\beta^{(0)}}\rangle
\langle \mathcal{D}^{{\alpha-}^{(0)}}_q\rangle
\langle{X}_i^{\alpha^{(0)}}\rangle\biggr)\\
&+2\pi\delta[\Delta'+\Delta]\biggl(-\frac{1}{2}
\underline{G^{\beta}_{i'j'}[\Delta']G^{\alpha}_{ij}[\Delta]
D^{\beta\alpha^{(0)}}_{j'j}}
+G_{ij}^{\alpha^+_q}[\Delta]G_{i'j'}^{\beta}[\Delta']
G_{\mathcal{D}^-_qk'}^{\beta}[\Delta]
D^{\beta\beta^{(0)}}_{j'k'}\langle X_j^{\alpha^{(0)}}\rangle
\biggr.\\
&\phantom{+2\pi\delta[\Delta'+\Delta]\biggl.-\frac{1}{2}}
+G_{i'j'}^{\beta^+_q}[\Delta']G_{\mathcal{D}^-_qk}^{\alpha}[\Delta']
G_{ij}^{\alpha}[\Delta]D^{\alpha\alpha^{(0)}}_{kj}\langle
X_{j'}^{\beta^{(0)}}\rangle\biggr).
\end{aligned}
\end{equation}
The corresponding stationary solution then reads:
\begin{equation}
\begin{aligned}
\langle{X}_{i'}^{\beta}{X}_i^{\alpha}\rangle^{(g)}&=
G_{ij}^{\alpha^+_q}[0]\langle X_j^{\alpha^{(0)}}\rangle
\langle{X}_{i'}^{\beta^{(0)}}\rangle
\langle\mathcal{D}^{{\beta-}^{(0)}}_q\rangle
+G_{i'j'}^{\beta^+_q}[0] \langle X_{j'}^{\beta^{(0)}}\rangle
\langle \mathcal{D}^{{\alpha-}^{(0)}}_q\rangle
\langle{X}_i^{\alpha^{(0)}}\rangle\\
&+\frac{1}{2\pi}\int d\Delta\biggl(-\frac{1}{2}
\underline{G^{\beta}_{i'j'}[-\Delta]G^{\alpha}_{ij}[\Delta]
D^{\beta\alpha^{(0)}}_{j'j}}
+G_{ij}^{\alpha^+_q}[\Delta]G_{i'j'}^{\beta}[-\Delta]
G_{\mathcal{D}^-_qk'}^{\beta}[\Delta]
D^{\beta\beta^{(0)}}_{j'k'}\langle X_j^{\alpha^{(0)}}\rangle
\biggr.\\
&\phantom{+\frac{1}{2\pi}\int d\Delta\biggl.-\frac{1}{2}}
+G_{i'j'}^{\beta^+_q}[-\Delta]G_{\mathcal{D}^-_qk}^{\alpha}[-\Delta]
G_{ij}^{\alpha}[\Delta]D^{\alpha\alpha^{(0)}}_{kj}\langle
X_{j'}^{\beta^{(0)}}\rangle\biggr).
\end{aligned}
\end{equation}

All quantities above only depend on the stationary values without
coupling between the atoms and thus can be calculated from the
single atom solutions. Furthermore, the integration over $\Delta$
can be performed either numerically or analytically by the theorem
of residues once the poles of $G$ (\emph{i.e.} the complex
eigenvalues of $M$) are known. Because of causality, they
all lie in the lower-half of the complex plane. In
practice, we have checked that we effectively recover, from the
preceding expressions, the results obtained from the full OB
equations. In particular, the contribution of the correlations of
the quantum vacuum
  fluctuations evaluated at the two atom positions (the underlined term) is
  essential to get the correct results.

The same kind of expressions can be derived for $g\bar{g}$ terms,
but they are slightly more complicated, since they explicitly
involve three-body correlation functions, more precisely terms
like:
\begin{equation}
\left\{\begin{aligned}
&G_{ij}^{\alpha^+_q}[\Delta]\left\langle{X}_{i'}^{\beta^{(0)}}[\Delta']
\bigl(X_j^{\alpha^{(0)}}\otimes
\mathcal{D}^{{\beta-}^{(0)}}_q\bigr)[\Delta]\right\rangle^{(\bar{g})}\\
&G_{ij}^{\alpha^+_q}[\Delta]\left\langle {X}_{i'}^{\beta}[\Delta']
\left(G^{\alpha^-_p}_{jj'}
\left(\mathcal{D}^{{\beta+}^{(0)}}_p\otimes
X_{j'}^{\alpha^{(0)}}\right)\otimes
\mathcal{D}^{{\beta-}^{(0)}}_q\right)[\Delta]\right\rangle^{(0)}\\
\end{aligned}\right.,
\end{equation}
which require the calculation of three-points Langevin force
correlation functions like:
\begin{equation}
\left\{\begin{aligned}
&G_{ij}^{\alpha^+_q}[\Delta]G^{\beta}_{i'j'}[\Delta']
\frac{1}{2\pi}\iint
d\Delta_1d\Delta_2\delta[\Delta_1+\Delta_2-\Delta]
G_{jk}^{\alpha}[\Delta_1]G_{\mathcal{D}^-_qk'}^{\beta}[\Delta_2]
\left\langle F^{\beta}_{j'}[\Delta']F^{\alpha}_{k}[\Delta_1]
F^{\beta}_{k'}[\Delta_2]\right\rangle^{(\bar{g})}\\
&G_{ij}^{\alpha^+_q}[\Delta]G^{\beta}_{i'k'}[\Delta']
\frac{1}{2\pi}\iint
d\Delta_1d\Delta_2\delta[\Delta_1+\Delta_2-\Delta]
G^{\alpha^-_p}_{jj'}[\Delta_1]G_{\mathcal{D}^+_pk}^{\beta}[\Delta_1]
G_{\mathcal{D}^+_pk''}^{\beta}[\Delta_2] \left\langle
F^{\beta}_{k'}[\Delta']F^{\beta}_{k}[\Delta_1]
F^{\beta}_{k''}[\Delta_2]\right\rangle^{(0)}
\end{aligned}\right..
\end{equation}

These correlations functions are non-zero even if they involve an
odd number of Langevin forces, emphasizing that the statistical
properties of the vacuum field fluctuations are far from Gaussian.
Nevertheless, the explicit expressions of the above quantities can
be derived (see appendix~\ref{matdthreebody}). They lead to rather
complicated and tedious formulae for the  atomic correlation
functions at order $g\bar{g}$. From that, we get the corresponding
stationary expectations values. Again, we have checked that we
indeed recover the OB results.

\subsection{Incorporation of an effective medium}

Finally, and in sharp contrast to optical Bloch equations,
it is very easy to adapt all the preceding results to the
case of propagation in a medium with a frequency-dependent
complex susceptibility. Indeed, propagation is controlled by the complex
amplitude $g$ so that the field radiated by an atom at a distance
$R$ and at frequency $\Delta$ will be given by:
\begin{equation}
\Omega^+_q[\Delta]= i \, g \, \mathcal{P}^{\textbf{R}}_{qq'}
\mathcal{D}^-_{q'}[\Delta]
\exp{\left(-\frac12\frac{R}{\ell^{+}[\Delta]}\right)},
\end{equation}
where $\ell^+[\Delta]$ is the (complex) scattering mean-free path
satisfying the dilute regime condition $k|\ell^+[\Delta]|\gg 1$.
The real part of  $1/\ell^+[\Delta]$ describes the exponential
attenuation of the field during its propagation in the medium
while the imaginary part describes the additional dephasing
induced by the medium.
More complicated formulas, accounting for possible variations of
$\ell$ with position, birefringence effects, or even
nonlinearities in propagation, can be derived in the same spirit.
In all preceding equations, leading to the calculation of the
correlation functions, any occurrence of the dipole operators must
then simply be replaced by:
\begin{equation}
\mathcal{D}^{\mp} \quad \to \quad
\mathcal{D}^{\mp}\exp{\left(-\frac{R}{2\ell^{\pm}[\Delta]}\right)}
\end{equation}
while keeping the same "medium-free" coupling constant $g$. In
this way, the present approach can be easily extended to the
situation where the two atoms are embedded in a medium. In the
case of a nonlinear medium, this could lead to a self-consistent
set of nonlinear equations.

It is important to stress that accounting for the effective medium
is rather straightforward in this frequency-domain approach but is
a much more difficult task in the temporal-domain approach.
Indeed, one basic hypothesis for deducing OB equations from the
Langevin approach -- see section~\ref{equivalence_ob} -- is that
the light propagation time between the two atoms is much shorter
than any typical atomic timescale. When this condition is
fulfilled, it is possible to evaluate expectation values at equal
times for both atoms, producing the set of closed OB equations. In
the presence of a surrounding medium, propagation between the two
atoms is affected and this basic assumption may fail. If the
refraction index of the dilute medium is smoothly varying with
frequency, then the corresponding propagation term is also
smoothly varying with frequency and can be factored out. Thus,
except for the exponential attenuation, one may recover the OB
equations where equal times must be used for atoms 1 and 2. On the
contrary, if the propagation term has a complicated frequency
dependence, the problem cannot be simply reduced to OB equations.
It will rather involve operators evaluated at the other atom, but
\emph{at different times}, thus leading to a much more complicated
structure. This difficulty may even take place in a dilute medium
with refraction index close to unity. Indeed, the important
parameter is the time delay induced by the medium, itself related
to the \emph{derivative} of the index of refraction with respect
to frequency. If the medium is composed of atoms having sharp
resonances, the effective group velocity can be reduced by several
orders of magnitude, consequently increasing by the same amount
the propagation time between the two atoms. Around the atomic
resonance line, the typical propagation time delay induced by the
medium over one mean free path depends on the laser detuning but
is of the order of the atomic timescale for the internal dynamics,
namely $\Gamma^{-1}$~\cite{Labeyrie:radiation_trapping}. In this
case, only the full Langevin treatment developed in this paper can
properly account for the effect of the average atomic medium. Its
practical implementation calls for an investigation on its own and
is thus postponed to a future paper. We must also note that, if
the surrounding medium is composed of the same atoms than the
scatterers, it is not completely clear that propagation in the
medium can be described ``classically", \emph{i.e.} that the
correlation between the Langevin forces acting on the scatterers
and the Langevin forces acting on the medium can be safely
neglected.

For the rest of this paper, we will consider two isolated atoms in vacuum.

\section{Main results}

\subsection{Scattered field correlation functions in the CBS configuration}

In the case of a large number of atoms and for a given
configuration, the interference between all possible multiple
scattering paths gives rise to a speckle pattern. When averaging
the intensity scattered off the sample over all possible positions
of the atoms, one recovers the CBS phenomenon: the intensity
radiated in the direction opposite to the incident beam is up to
twice larger than the background intensity and gradually decreases
to the background value over an angular range $\Delta\theta$
scaling essentially as $(k\ell)^{-1}$, with $\ell$ the scattering
mean-free path. In the present case, the averaging procedure is
performed numerically by integrating over the relative positions
of the two atoms. As will be seen below, the far-field condition
$kR\gg 1$ allows for an \emph{a priori} selection of the dominant
terms contributing to the CBS signal.

The field radiated by the two atoms in the direction $\textbf{n}$
at a distance $r\gg R\gg\lambda$, in the polarization channel
$\boldsymbol{\epsilon}^{\mathrm{out}}$ orthogonal to $\textbf{n}$
($\boldsymbol{\epsilon}^{\mathrm{out}}\cdot \textbf{n} =0$), is
given by:
\begin{equation}
\Omega^{+}_{\mathrm{out}}[\textbf{n},\Delta]=-\frac{3}{2}\Gamma
\epsilon^{\mathrm{out}}_q
\left(\mathcal{D}^{1-}_{q}[\Delta]e^{-ik\textbf{n}\cdot\textbf{R}_1}
+\mathcal{D}^{2-}_{q}[\Delta]e^{-ik\textbf{n}\cdot\textbf{R}_2}\right)
\frac{e^{ikr}}{kr},
\end{equation}
so that the field correlation function in this channel reads:
\begin{multline}
\label{field1}
\langle\Omega^{-}_{\mathrm{out}}[\textbf{n},\Delta']
\Omega^{+}_{\mathrm{out}}[\textbf{n},\Delta]\rangle=
\big(\frac{3\Gamma}{2kr}\big)^2 \,
{\epsilon}^{\mathrm{out}}_q\epsilon^{\mathrm{out}}_{p} \biggl\{
\langle\mathcal{D}^{1+}_{p}[\Delta']\mathcal{D}^{1-}_{q}[\Delta]\rangle+
\langle\mathcal{D}^{2+}_{p}[\Delta']
\mathcal{D}^{2-}_{q}[\Delta]\rangle\biggr.\\
\biggl.+e^{ik\textbf{n}\cdot\textbf{R}}
\langle\mathcal{D}^{2+}_{p}[\Delta']\mathcal{D}^{1-}_{q}[\Delta]\rangle+
e^{-ik\textbf{n}\cdot\textbf{R}}
\langle\mathcal{D}^{1+}_{p}[\Delta']\mathcal{D}^{2-}_{q}[\Delta]\rangle
\biggr\}.
\end{multline}

The CBS effect occurs when the total phase in the interference
terms in the preceding expression becomes independent of the
positions of the atom. This phase accumulates during the
propagation of the incident laser beam to the atoms and during the
propagation of the radiated field between the two atoms. The phase
factor due to the incoming laser beam (a plane wave with wave
number $\textbf{k}_L = k \, \textbf{n}_L$) can be explicitly
factorized out of the atomic operators as follows:
\begin{equation}
\tilde{\mathcal{D}}^{\alpha\pm}_q =\mathcal{D}^{\alpha\pm}_q\,
e^{\pm i\textbf{k}_L\cdot\textbf{R}_{\alpha}}.
\end{equation}
The other components of $\tilde{X}$, cf. Eq.~(\ref{def_X}),
are populations and not affected by this phase factor.
In the single atom case, the expectation values of the hereby defined
operators
$\tilde{\mathcal{D}}^{\alpha\pm}_q$ are independent of the
positions of the atoms. Defining $\phi=\textbf{k}_L\cdot\textbf{R}$
and
\begin{equation}
g_1=ge^{i\phi}\qquad g_2=ge^{-i\phi} ,
\end{equation}
the Langevin equations~\eqref{langevin2} then become:
\begin{equation}
\label{langevin3}
\tilde{\mathbf{X}}^{\alpha}[\Delta]=\tilde{\mathbf{X}}^{\alpha^{(0)}}[\Delta]+
g_{\alpha}\tilde{G}^{\alpha_q^+}[\Delta]\left(\tilde{\mathbf{X}}^{\alpha}\otimes
\tilde{\mathcal{D}}^{\beta-}_{q}\right)[\Delta]+
\bar{g}_{\alpha}\tilde{G}^{\alpha_q^-}[\Delta]
\left(\tilde{\mathcal{D}}^{\beta+}_q
\otimes\tilde{\mathbf{X}}^{\alpha}\right)[\Delta],
\end{equation}

In the preceding equation, the Green's functions $\tilde{G}$ are
now independent of the position of the atoms, so that the phase
information due to the incident laser beam is entirely contained
in the coefficients $g_{\alpha}$.

Frequency correlation functions of the Langevin forces,
eq.~\eqref{corrfafab}, must also be modified accordingly:
\begin{equation}
\langle \tilde{F}_{i'}^{\beta}[\Delta']\tilde{F}_i^{\alpha}[\Delta]\rangle=
-\frac{1}{2}\biggl(g_{\beta}+\bar{g}_{\alpha}\biggr)
2\pi\delta[\Delta'+\Delta]\tilde{D}^{\beta\alpha}_{i'i}.
\end{equation}
Dropping for simplicity, the tilde notation, the field correlation
function~\eqref{field1}, in the backward direction
$\textbf{n}=-\textbf{n}_L$, becomes:
\begin{multline}
\label{field2}
\langle\Omega^{-}_{\mathrm{out}}[-\textbf{n}_L,\Delta']
\Omega^{+}_{\mathrm{out}}[-\textbf{n}_L,\Delta]\rangle=
\left(\frac{\Gamma}{kr}\right)^2 \,
{\epsilon}^{\mathrm{out}}_q\epsilon^{\mathrm{out}}_{p}\biggl\{
\langle\mathcal{D}^{1+}_{p}[\Delta']\mathcal{D}^{1-}_{q}[\Delta]\rangle+
\langle\mathcal{D}^{2+}_{p}[\Delta']\mathcal{D}^{2-}_{q}[\Delta]\rangle
\biggr.\\
\biggl.+e^{-2i\phi}
\langle\mathcal{D}^{2+}_{p}[\Delta']\mathcal{D}^{1-}_{q}[\Delta]\rangle+
e^{2i\phi}
\langle\mathcal{D}^{1+}_{p}[\Delta']\mathcal{D}^{2-}_{q}[\Delta]\rangle
\biggr\}.
\end{multline}
The configuration average is then performed in two steps. Since we
are working in the limit $kR\gg 1$, the first one is to keep only
terms with a total phase independent of $kR$. In the power
expansion with respect to the four parameters $g_1$, $g_2$,
$\bar{g}_1$ and $\bar{g}_2$, this simply amounts to keep terms
with even powers of $g_{\alpha}\bar{g}_{\alpha'}$. This obviously
cancels any $\phi$ dependence. More precisely, the
field correlation function in the backward direction, beside
the trivial zeroth order (in $g$) term, is given by:
 \begin{equation}
\label{fieldmoy}
\begin{aligned}
\langle\Omega^{-}_{\mathrm{out}}[-\textbf{n}_L,\Delta']
\Omega^{+}_{\mathrm{out}}[-\textbf{n}_L,\Delta]\rangle^{(2)}&=
\left(\frac{\Gamma}{kr}\right)^2 \,
{\epsilon}^{\mathrm{out}}_q\epsilon^{\mathrm{out}}_{p}\biggl\{
\langle\mathcal{D}^{1+}_{p}[\Delta']\mathcal{D}^{1-}_{q}[\Delta]
\rangle^{(g_1\bar{g}_1)}+ \langle\mathcal{D}^{2+}_{p}[\Delta']
\mathcal{D}^{2-}_{q}[\Delta]\rangle^{(g_2\bar{g}_2)}\biggr.\\
&\phantom{\frac{1}{k^2R^2}\frac{9}{4}\Gamma^2
{\epsilon}^{\mathrm{out}}_q\epsilon^{\mathrm{out}}_{p}}\biggl.+
\langle\mathcal{D}^{2+}_{p}[\Delta']\mathcal{D}^{1-}_{q}[\Delta]
\rangle^{(g_1\bar{g}_2)}+
\langle\mathcal{D}^{1+}_{p}[\Delta']\mathcal{D}^{2-}_{q}[\Delta]
\rangle^{(g_2\bar{g}_1)}
\biggr\}\\
&= \left(\frac{\Gamma}{kr}\right)^2 \,
\big(L[\Delta',\Delta]+C[\Delta',\Delta]\big).
\end{aligned}
\end{equation}

The preceding field correlation function still depends on the
relative orientation of the atoms through the projector
$\mathcal{P}^{\textbf{R}}$, so that, in a second step, an
additional average over $\textbf{R}$ must be performed. In the
preceding equation, the first two terms correspond to the usual
``ladder'' terms $L[\Delta',\Delta]$ (they are actually
independent of the direction of observation), whereas the two
other terms correspond to the usual ``maximally crossed'' terms
$C[\Delta',\Delta]$:
\begin{equation}
\begin{aligned}
L[\Delta',\Delta] =\frac{9}{4}
{\epsilon}^{\mathrm{out}}_q\epsilon^{\mathrm{out}}_{p} \biggl\{
\langle\mathcal{D}^{1+}_{p}[\Delta']\mathcal{D}^{1-}_{q}[\Delta]
\rangle^{(g_1\bar{g}_1)}+ \langle\mathcal{D}^{2+}_{p}[\Delta']
\mathcal{D}^{2-}_{q}[\Delta]\rangle^{(g_2\bar{g}_2)}\biggl\} \\
C[\Delta',\Delta] =\frac{9}{4}
{\epsilon}^{\mathrm{out}}_q\epsilon^{\mathrm{out}}_{p}
\biggl\{\langle\mathcal{D}^{2+}_{p}[\Delta']\mathcal{D}^{1-}_{q}[\Delta]
\rangle^{(g_1\bar{g}_2)}+
\langle\mathcal{D}^{1+}_{p}[\Delta']\mathcal{D}^{2-}_{q}[\Delta]
\rangle^{(g_2\bar{g}_1)}\biggl\}
\end{aligned}
\end{equation}

\subsection{CBS enhancement factor}

In the case of linear scatterers, the CBS enhancement factor
achieves its maximal value 2 (recall that the CBS phenomenon is an
incoherent sum of two-wave interference patterns all starting with
a bright fringe at exact backscattering) if the single scattering
contribution can be removed from the total signal and provided
reciprocity holds. This is the case for scatterers with spherical
symmetry in the so-called polarization preserving channel $h
\parallel h$~\cite{BvTMaynard}.

In this polarization channel, we have calculated the relevant
quantities for an evaluation of the CBS enhancement factor
\emph{when no frequency filtering of the outgoing signal is made}.
We have thus derived the elastic and inelastic ladder terms and
the elastic and inelastic crossed terms, together with their
corresponding frequency spectra, for different values of the
on-resonance saturation parameter $s_0=2|\Omega_L|^2/\Gamma^2$.
This parameter measures the intensity strength of the incident
laser beam in units of the natural atomic transition line width
$\Gamma$, \emph{i.e.} its compares the on-resonance transition
rate induced by the laser to the atomic spontaneous emission rate.
For a detuned laser beam, the saturation parameter is $s(\delta)$
and is defined as:
\begin{equation}
s(\delta) = \frac{s_0}{1+(2\delta/\Gamma)^2}
\end{equation}

In the following, different values of the laser detuning have also
been considered:
\begin{equation*}
\begin{array}{ll}
(a)\quad \delta=0,\, s = s_0 = 0.02 &\quad
(b)\quad \delta=0,\, s = s_0 = 2.00 \\
(c)\quad \delta=5\Gamma, \,  s_0=2.00, \, s=0.02
&\quad(d)\quad \delta=0,\, s = s_0 = 50.0
\qquad\qquad
\end{array}.
\end{equation*}

The ladder and crossed terms~\eqref{fieldmoy} are separated into
their elastic and inelastic parts according to:
\begin{equation}
\label{laddcrossplot}
\begin{aligned}
L[\Delta',\Delta]&=2\pi\delta(\Delta+\Delta')
\, \bigl\{2\pi\delta(\Delta) \, L_{\mathrm{el}}+L_{\mathrm{inel}}(\Delta)\bigr\}\\
C[\Delta',\Delta]&=2\pi\delta(\Delta+\Delta') \,
\bigl\{2\pi\delta(\Delta) \,
C_{\mathrm{el}}+C_{\mathrm{inel}}(\Delta)\bigr\}
\end{aligned}
\end{equation}

\begin{figure}[h]
\includegraphics[angle=-90,width=15cm]{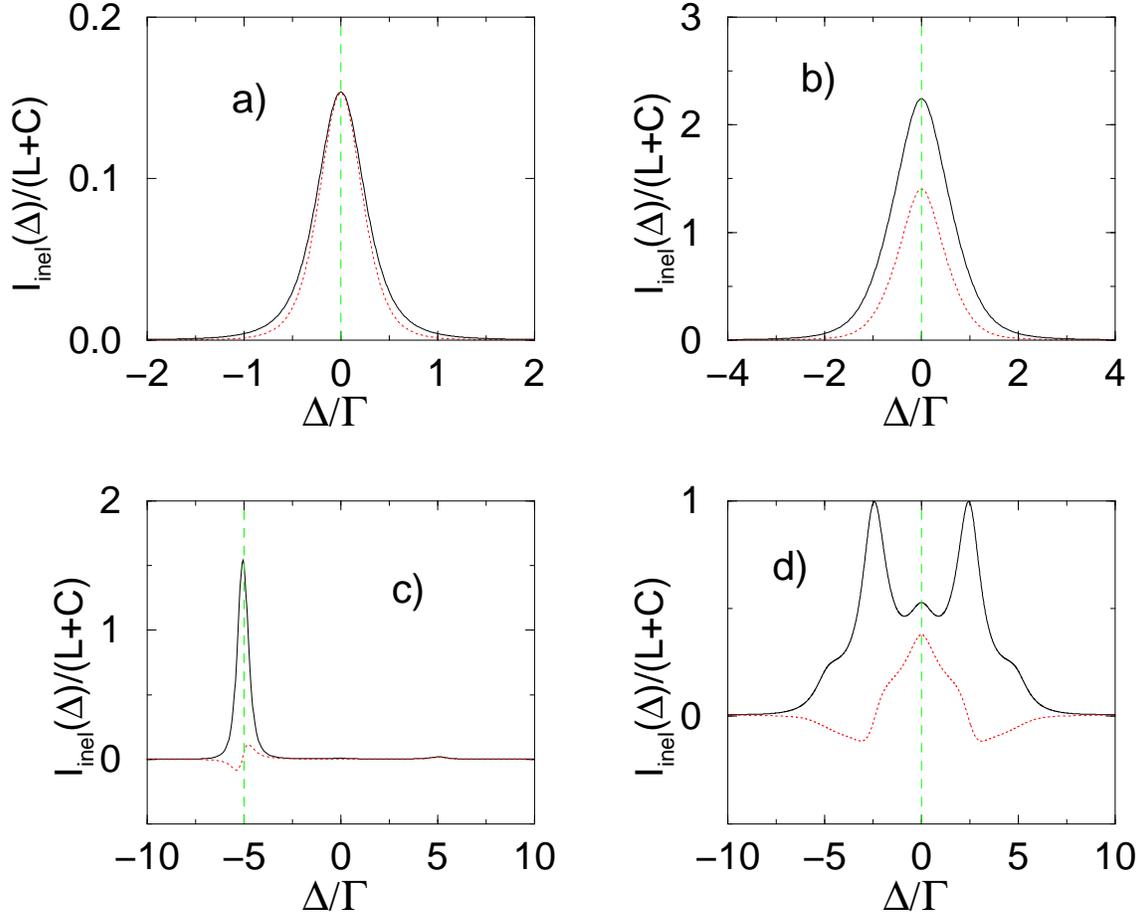}
\caption{\label{specfig} Backscattered light spectrum in the
  helicity-preserving polarization channel $h\parallel h$. The solid lines
  represent
  the ladder term (average background intensity value) and the dotted lines represent the crossed
   (interference) term. For both terms, the plotted values
  corresponds to
$I_{\mathrm{inel}}(\Delta)/(C^{\mathrm{tot}}+L^{\mathrm{tot}})$,
 see Eq.\eqref{laddcrossplot},
where $C^{\mathrm{tot}}+L^{\mathrm{tot}}$ is the total (elastic
\emph{plus} inelastic) intensity
  scattered in the backward direction.
  The
  vertical dashed lines indicate the atomic
  transition frequency. $\Delta$ corresponds to the scattered light angular frequency change
  with respect to the initial laser angular frequency ($\Delta=0$ means thus that light is radiated at $\omega_L$).
  Graph ($a$) corresponds to an on-resonance saturation parameter $s_0=0.02$ and
  a laser detuning $\delta=0$ ; Graph ($b$) to $(s_0=2,\delta=0)$ ; Graph
  ($c$) to $(s_0=2,\delta=5\Gamma)$ and Graph ($d$) to $(s_0=50,\delta=0)$.
  At low $s_0$, the inelastic contribution to the
  total intensity is small and the ladder intensity is almost equal to
  the crossed one. For a larger saturation parameter, firstly the inelastic contribution becomes
  comparable to
 the elastic one and secondly, the crossed term becomes smaller than
the
  ladder one.
For a nonzero detuning, see graph ($c$), one clearly
  observes an asymmetry in the
  inelastic spectrum, which reflects the fact that the scattering
  cross-section
  of the atomic transition is maximal for resonant light: the
  symmetric inelastic spectrum emitted by a single atom is filtered out
  when scattered by the other one. At very large saturation (d),
  the structure of the radiated spectrum becomes rather complicated.
}
  \end{figure}

The corresponding inelastic spectra $L_{\mathrm{inel}}(\Delta)$
and $C_{\mathrm{inel}}(\Delta)$ are displayed in
figure~\ref{specfig}. For a sufficiently low saturation parameter
$s_0$, the inelastic contribution to the total intensity is small
and the ladder intensity is almost equal to the crossed one (see
graph~\ref{specfig}$a$). For larger saturation parameters (see
graphs~\ref{specfig}$b$ and \ref{specfig}$d$), there are two
effects : first, the inelastic contribution becomes comparable to
the elastic one and second, the crossed term is smaller than the
ladder one. For a nonzero detuning (see graph \ref{specfig}$c$),
one clearly observes an asymmetry in the inelastic spectrum, which
reflects that the scattering cross-section of the atomic
transition is maximal for resonant light (indicated by the
vertical dashed line): the symmetric inelastic spectrum emitted
by a single atom is filtered out when scattered by the other one.
We also observe that the crossed spectrum is much more reduced
than the ladder term, highlighting the non-linear effects in the
quantum correlations between the two atoms. Finally, for much
larger saturation parameters (see graph~\ref{specfig}$d$),
the scattered light almost entirely originates from the inelastic spectrum,
like for a single atom. However, contrary to the single atom case (for
which the scattered intensity reaches a constant value), the
total intensity scattered by the two atoms decreases  when increasing 
the incoming intensity. Indeed, since the atomic transitions become fully
saturated, the nonlinear scattering cross-section of each atom is
decreasing, resulting in a smaller total intensity scattered by
the two atoms compared to the one scattered by a single atom. 

The CBS enhancement factor $\eta$ is defined as the peak to
background ratio. It thus reads:
\begin{equation}
\eta = 1 + \frac{C^{\mathrm{tot}}}{L^{\mathrm{tot}}}
\end{equation}
with:
\begin{equation}
\begin{aligned}
L^{\mathrm{tot}}&=L_{\mathrm{el}}+L_{\mathrm{inel}}^{\mathrm{tot}}=
L_{\mathrm{el}}+ \int \frac{d\Delta}{2\pi} \,L_{\mathrm{inel}}(\Delta)\\
C^{\mathrm{tot}}&=C_{\mathrm{el}}+C_{\mathrm{inel}}^{\mathrm{tot}}=
C_{\mathrm{el}}+ \int
\frac{d\Delta}{2\pi}\,C_{\mathrm{inel}}(\Delta)
\end{aligned}
\end{equation}

If the CBS phenomenon is reducible to a  two-wave
  interference, as it is the case here, then the enhancement factor $\eta$
  is simply related to the degree of coherence $\gamma$ of the scattered light \cite{coherence}.
 If the single scattering contribution can be removed from the
detected signal, and this is the case in the $h
\parallel h$ channel, one has simply $\eta=1+\gamma$ and consequently $\gamma =
C^{\mathrm{tot}}/L^{\mathrm{tot}}$. The maximal value for $\eta$
is 2, meaning that full coherence $\gamma=1$ is maintained for the
scattered field since then $C^{\mathrm{tot}}=L^{\mathrm{tot}}$.
If all interference effects disappear, meaning
$C^{\mathrm{tot}}=0$, $\eta$ reaches its minimal value 1 and
correspondingly coherence is fully lost $\gamma=0$. Furthermore,
one can show that in the $h \parallel h$ polarization channel,
$L_{\mathrm{el}} = C_{\mathrm{el}}$~\cite{prl94SMB}. Consequently,
as soon as $C_{\mathrm{inel}}^{\mathrm{tot}} <
L_{\mathrm{inel}}^{\mathrm{tot}}$ in this channel, the
coherence of the scattered light field is partially destroyed, since
then $\eta < 2$ and $\gamma <1$.

\begin{table}[ht]
\caption{\label{table1} Ladder (average background) and crossed
(interference) terms, see Eq.\eqref{laddcrossplot}, contributing
to the light scattered in the backward direction in the
helicity-preserving polarization channel $h\parallel h$. The given
values are relative to the incoming saturation parameter $s$. At
low $s_0$, the inelastic contributions are small and almost equal.
Thus $C^{\mathrm{tot}} \approx L^{\mathrm{tot}}$ and the maximum
enhancement factor 2 of the linear case is thus recovered, meaning
that full coherence $\gamma=1$ is maintained. At larger $s_0$,
elastic and inelastic terms become comparable. For very large
$s_0$, the contributions from the elastic terms vanish, like in
the single atom case. The inelastic contributions are also
decreasing, reflecting the fact that the probability for the light
to be scattered by a saturated atom becomes smaller with
increasing saturation. Furthermore, the inelastic crossed term is
\emph{always} smaller than the inelastic ladder one. This is a
signature of a coherence loss $\gamma <1$ induced by the quantum
vacuum fluctuations. However, the ratio
$C_{\mathrm{inel}}^{\mathrm{tot}}/L_{\mathrm{inel}}^{\mathrm{tot}}$
does not go to zero as $s_0 \to \infty$ but reaches the limit
value 0.096 (for $\delta=0$). Also, contrary to the single atom case, the
properties of the scattered light are not solely determined by
the saturation parameter $s$, but additionally depend on the
detuning $\delta,$ as exemplified by cases (a) and (c),
highlighting the role of the inelastic processes.}
\begin{ruledtabular}
\begin{tabular}{rd@{}d@{}d@{}d@{}}
 & \multicolumn{1}{r}{$(a)\,s=s_0=0.02, \delta=0$}
&  \multicolumn{1}{r}{$(b)\,s=s_0=2.00, \delta=0$}
& \multicolumn{1}{r}{$(c)\,s=0.02, s_0=2.00, \delta=5\Gamma$}
& \multicolumn{1}{r}{$(d)\,s=s_0=50.0, \delta=0$} \\
\multicolumn{1}{l}{$L_{\mathrm{el}}$} & 0.624 &0.833E-02 &
0.612 & 0.998E-07\\
\multicolumn{1}{l}{$L_{\mathrm{inel}}^{\mathrm{tot}}$} & 0.220E-01
&0.573E-01 & 0.328 & 0.487E-03 \\
\multicolumn{1}{l}{$L^{\mathrm{tot}}$} & 0.646 &0.656E-01 &
0.946 & 0.487E-03 \\
\multicolumn{1}{l}{$C_{\mathrm{el}}$} & 0.624 & 0.833E-02 &
0.612 & 0.998E-07\\
\multicolumn{1}{l}{$C_{\mathrm{inel}}^{\mathrm{tot}}$} & 0.188E-01 &
0.295E-01 & 0.157E-01 & 0.466E-04 \\
\multicolumn{1}{l}{$C^{\mathrm{tot}}$} & 0.642 & 0.378E-01 &
0.634 & 0.467E-04\\
\multicolumn{1}{l}{$\eta=1+\gamma$} & 1.994 & 1.576 &
1.670 & 1.096\\
\end{tabular}
\end{ruledtabular}
\end{table}

Our results are summarized in table~\ref{table1}. At low
saturation parameter $s_0$, $\eta$ reaches its maximal value 2 and
$\gamma=1$. This is so because the ladder and crossed inelastic
components are almost equal as evidenced in \ref{specfig}$a$.
Increasing $s_0$ reduces further
$C_{\mathrm{inel}}^{\mathrm{tot}}$ with respect to
$L_{\mathrm{inel}}^{\mathrm{tot}}$, thus decreasing $\eta$ and
$\gamma$.  In the strongly saturated regime, one thus expects
$\gamma$ to decrease. However, there is no reason for the ratio
$C_{\mathrm{inel}}^{\mathrm{tot}}/
L_{\mathrm{inel}}^{\mathrm{tot}}$ to tend to zero as $s_0\to
\infty.$ It rather tends to a finite value, which depends on the
detuning, in agreement with the results published in
\cite{prl94SMB}. Furthermore, keeping $s_0$ fixed and decreasing
the saturation parameter $s$, situation $(c)$, $\eta$ increases,
as expected, but to a value which strongly depends on $s_0$. In
other words, contrary to the single atom case, the properties of
the scattered light, are not only determined by the saturation
parameter $s$~\cite{pra70WGDM}. Indeed, in both situations $(a)$
and $(c)$, $s$ has the same (small) value, but the enhancement
factor strongly differs, mainly because the inelastic ladder term
has increased. This highlights the crucial role of the inelastic
processes and of the rather complicated quantum correlations
between the two atoms.

This is not however the full story. Depending on the $s$ and
$\delta$ parameters, a rich variety of situations can be observed,
with various physical interpretations. These are beyond the scope
of this paper, which instead concentrates on the basic ingredients
of the quantum Langevin approach and will be published elsewhere.

\subsection{Linear response model}

Some insight on the relative behavior of
$C_{\mathrm{inel}}(\Delta)$ and $L_{\mathrm{inel}}(\Delta)$  can
be found by comparing the respective formulae from which these
quantities are extracted:
\begin{multline}
\label{corrcbscross} \left\langle{X}_{i'}^{\beta}[\Delta']
{X}_i^{\alpha}[\Delta]\right\rangle^{(\bar{g}_{\beta}g_{\alpha})}=
g_{\alpha}\left\langle{X}_{i'}^{\beta^{(0)}}[\Delta']
G_{ij}^{\alpha^+_q}[\Delta] \bigl(X_j^{\alpha^{(0)}}\otimes
\mathcal{D}^{{\beta-}^{(0)}}_q\bigr)[\Delta]
\right\rangle^{(\bar{g}_{\beta})} -\bar{g}_{\beta} \left\langle
G_{i'j'}^{\beta^-_q}[\Delta']
\bigl(\mathcal{D}^{{\alpha+}^{(0)}}_q\otimes
X_{j'}^{\beta^{(0)}}\bigr)[\Delta']
{X}_i^{\alpha^{(0)}}[\Delta]\right\rangle^{(g_{{\alpha}})}\\
-g_{\alpha}\bar{g}_{\beta}\biggl\{
\left\langle{X}_{i'}^{\beta}[\Delta']
G_{ij}^{\alpha^+_q}[\Delta]\left(X_j^{\alpha^{(0)}} \otimes
G^{\beta^-_p}_{\mathcal{D}_q^-j'}
\left(\mathcal{D}^{{\alpha+}^{(0)}}_p
\otimes X^{\beta^{(0)}}_{j'}\right)\right)[\Delta]\right\rangle^{(0)}\\
+\left\langle G_{i'j'}^{\beta^-_q}[\Delta']
\left(G^{\alpha^+_p}_{\mathcal{D}_q^-j}
\left(X_{j}^{\alpha^{(0)}}\otimes
\mathcal{D}^{{\beta-}^{(0)}}_p\right)\otimes
X_{j'}^{\beta^{(0)}}\right)[\Delta']
{X}_i^{\alpha^{(0)}}[\Delta]\right\rangle^{(0)}\\
+\biggl.\left\langle\biggl[G_{i'j'}^{\beta^-_p}[\Delta']
\bigl(\mathcal{D}^{{\alpha+}^{(0)}}_p\otimes
X_{j'}^{\beta^{(0)}}\bigr)[\Delta']\biggr]
\biggl[G_{ij}^{\alpha^+_q}[\Delta] \bigl(X_j^{\alpha^{(0)}}\otimes
\mathcal{D}^{{\beta-}^{(0)}}_q\bigr)[\Delta]
\biggr]\right\rangle^{(0)}\biggr\}
\end{multline}
and
\begin{multline}
\label{corrcbsladder}
\left\langle{X}_{i'}^{\alpha}[\Delta']
{X}_i^{\alpha}[\Delta]\right\rangle^{(\bar{g}_{\alpha}g_{\alpha})}=
\left\langle{X}_{i'}^{\alpha^{(0)}}[\Delta']{X}_i^{\alpha^{(0)}}[\Delta]
\right\rangle^{(\bar{g}_{\alpha}g_{\alpha})}\\
+g_{\alpha}\biggl\{\left\langle{X}_{i'}^{\alpha^{(0)}}[\Delta']
G_{ij}^{\alpha^+_q}[\Delta] \bigl(X_j^{\alpha^{(0)}}\otimes
\mathcal{D}^{{\beta-}^{(0)}}_q\bigr)[\Delta]
\right\rangle^{(\bar{g}_{\alpha})} +\left\langle
G_{i'j'}^{\alpha^+_q}[\Delta'] \bigl(X_{j'}^{\alpha^{(0)}}\otimes
\mathcal{D}^{{\beta-}^{(0)}}_q\bigr)[\Delta']
{X}_i^{\alpha^{(0)}}[\Delta]\right\rangle^{(\bar{g}_{\alpha})}
\biggr\}\\
-\bar{g}_{\alpha}\biggl\{\left\langle{X}_{i'}^{\alpha^{(0)}}[\Delta']
G_{ij}^{\alpha^-_q}[\Delta]
\bigl(\mathcal{D}^{{\beta+}^{(0)}}_q\otimes
X_j^{\alpha^{(0)}}\bigr)[\Delta]\right\rangle^{(g_{\alpha})}
+\left\langle G_{i'j'}^{\alpha^-_q}[\Delta']
\bigl(\mathcal{D}^{{\beta+}^{(0)}}_q\otimes
X_{j'}^{\alpha^{(0)}}\bigr)[\Delta']
{X}_i^{\alpha^{(0)}}[\Delta]\right\rangle^{(g_{\alpha})}
\biggr\}\\
-\bar{g}_{\alpha}g_{\alpha}\biggl\{\left\langle{X}_{i'}^{\alpha}[\Delta']
G_{ij}^{\alpha^+_q}[\Delta]\left(G^{\alpha^-_p}_{jj'}
\left(\mathcal{D}^{{\beta+}^{(0)}}_p\otimes
X_{j'}^{\alpha^{(0)}}\right)\otimes
\mathcal{D}^{{\beta-}^{(0)}}_q\right)[\Delta]\right\rangle^{(0)}
\biggr.\\
+\left\langle{X}_{i'}^{\alpha}[\Delta']G_{ij}^{\alpha^-_q}[\Delta]
\left(\mathcal{D}^{{\beta+}^{(0)}}_q\otimes
G^{\alpha^+_p}_{jj'}\left(X_{j'}^{\alpha^{(0)}}\otimes
\mathcal{D}^{{\beta-}^{(0)}}_p\right)\right)[\Delta]
\right\rangle^{(0)}\\
+\left\langle
G_{i'j'}^{\alpha^+_q}[\Delta']\left(G^{\alpha^-_p}_{j'j}
\left(\mathcal{D}^{{\beta+}^{(0)}}_p\otimes
X_{j}^{\alpha^{(0)}}\right)\otimes
\mathcal{D}^{{\beta-}^{(0)}}_q\right)[\Delta']{
X}_i^{\alpha^{(0)}}[\Delta]\right\rangle^{(0)}\\
+\left\langle G_{i'j'}^{\alpha^-_q}[\Delta']
\left(\mathcal{D}^{{\beta+}^{(0)}}_q\otimes
G^{\alpha^+_p}_{j'j}\left(X_{j}^{\alpha^{(0)}}\otimes
\mathcal{D}^{{\beta-}^{(0)}}_p\right)\right)[\Delta']
{X}_i^{\alpha^{(0)}}[\Delta]\right\rangle^{(0)} \\
+\left\langle\biggl[G_{i'j'}^{\alpha^+_p}[\Delta']
\bigl(X_{j'}^{\alpha^{(0)}}\otimes
\mathcal{D}^{{\beta-}^{(0)}}_p\bigr)[\Delta']\biggr]
\biggl[G_{ij}^{\alpha^-_q}[\Delta]
\bigl(\mathcal{D}^{{\beta+}^{(0)}}_q\otimes
X_j^{\alpha^{(0)}}\bigr)[\Delta]\biggr]\right\rangle^{(0)}\\
+\biggl.\left\langle\biggl[G_{i'j'}^{\alpha^-_p}[\Delta']
\bigl(\mathcal{D}^{{\beta+}^{(0)}}_p\otimes
X_{j'}^{{\alpha}^{(0)}}\bigr)[\Delta']\biggr]
\biggl[G_{ij}^{\alpha^+_q}[\Delta] \bigl(X_j^{\alpha^{(0)}}\otimes
\mathcal{D}^{{\beta-}^{(0)}}_q\bigr)[\Delta]
\biggr]\right\rangle^{(0)}\biggr\}.
\end{multline}

There are twice as many terms contributing to the ladder
terms as to the crossed terms. A rather simple explanation of this
fact is
borrowed from the usual linear response theory. Indeed, each atom
is exposed to two fields : the incoming monochromatic field
(angular frequency $\omega_L$, wave vector $\textbf{k}_L$) and the
field scattered by the other atom (angular frequency $\omega_L +
\Delta$, wave vector $\textbf{k}_p$). In the far-field regime $R
\gg \lambda$, the incoming field is more intense than the
scattered field. It thus plays the role of a pump beam with
angular Rabi frequency $\Omega_L$, while the second weaker field
plays the role of a probe beam with angular Rabi frequency
$\Omega_p$. In this case, the response of each atom is simply
described by its nonlinear susceptibility~\cite{Cohenrouge,boyd}.
More precisely, forgetting about polarization effects, we have:
\begin{equation}
\begin{aligned}
\delta\mathcal{D}^+[\Delta]&=
e^{-i(2\textbf{k}_L-\textbf{k}_p)\cdot\textbf{R}_{\alpha}} \,
\chi_{\sst ++}[\Delta] \, \Omega_p^+
+e^{-i\textbf{k}_p\cdot\textbf{R}_{\alpha}} \, \chi_{\sst +-}[\Delta] \, \Omega_p^-\\
\delta\mathcal{D}^-[\Delta]&=
e^{i\textbf{k}_p\cdot\textbf{R}_{\alpha}} \, \chi_{\sst
-+}[\Delta] \, \Omega_p^+
+e^{i(2\textbf{k}_L-\textbf{k}_p)\cdot\textbf{R}_{\alpha}} \,
\chi_{\sst --}[\Delta] \, \Omega_p^-.
\end{aligned}
\end{equation}
where the phases due to the light fields have been explicitly
factorized.

As obviously seen, the two terms $\chi_{\sst +-}$ and $\chi_{\sst
-+}$ generate the forward propagation of the probe whereas the two
other terms $\chi_{\sst ++}$ and $\chi_{\sst --}$ can generate an
additional field in the direction $2\textbf{k}_L-\textbf{k}_p$
provided phase-matching conditions are fulfilled. This corresponds
to the usual forward four-wave mixing mechanism
(FFWM)~\cite{boyd,Cohenrouge}. If we now replace the probe field
by the field radiated by the other atom $\beta$, we get:
\begin{equation}
\begin{aligned}
\delta\mathcal{D}^+_{\beta\rightarrow\alpha}[\Delta]&=
\frac{1}{kR}\left\{ e^{-i(kR +
2\textbf{k}_L\cdot\textbf{R}_{\alpha} -
\textbf{k}_L\cdot\textbf{R}_{\beta})}\chi_{\sst ++}[\Delta]
\,\mathcal{D}^-_{\beta}
+e^{i(kR-\textbf{k}_L\cdot\textbf{R}_{\beta})}\chi_{\sst
+-}[\Delta]
\,\mathcal{D}^+_{\beta}\right\}\\
\delta\mathcal{D}^-_{\beta\rightarrow\alpha}[\Delta]&=
\frac{1}{kR}\left\{
e^{-i(kR-\textbf{k}_L\cdot\textbf{R}_{\beta})}\chi_{\sst
-+}[\Delta] \,\mathcal{D}^-_{\beta}
+e^{i(2\textbf{k}_L\cdot\textbf{R}_{\alpha}+kR
-\textbf{k}_L\cdot\textbf{R}_{\beta})}\chi_{\sst --}[\Delta]
\,\mathcal{D}^+_{\beta}\right\}.
\end{aligned}
\end{equation}

Hence the ladder and crossed contributions are given by (dropping
for sake of clarity any frequency dependence):
\begin{equation}
\begin{aligned}
C^{(2)}&\approx\delta\mathcal{D}^+_{\alpha\rightarrow\beta}
\delta\mathcal{D}^-_{\beta\rightarrow\alpha}
e^{i(-\textbf{k}_L\cdot\textbf{R}_{\beta}+\textbf{k}_L\cdot\textbf{R}_{\alpha})}\\
&\approx
e^{i(2\textbf{k}_L\cdot(\textbf{R}_{\alpha}-\textbf{R}_{\beta})-2kR)}
\chi_{\sst ++}\chi_{\sst
-+}\mathcal{D}^-_{\alpha}\mathcal{D}^-_{\beta}
+e^{4i\textbf{k}_L\cdot(\textbf{R}_{\alpha}-\textbf{R}_{\beta})}
\chi_{\sst ++}\chi_{\sst --}\mathcal{D}^-_{\alpha}\mathcal{D}^+_{\beta}\\
&\phantom{\approx} +\chi_{\sst +-}\chi_{\sst
-+}\mathcal{D}^+_{\alpha}\mathcal{D}^-_{\beta}
+e^{i(2\textbf{k}_L\cdot(\textbf{R}_{\alpha}-\textbf{R}_{\beta})+2kR)}
\chi_{\sst +-}\chi_{\sst --}\mathcal{D}^+_{\alpha}\mathcal{D}^+_{\beta}\\
\phantom{}\\
L^{(2)}&\approx\delta\mathcal{D}^+_{\beta\rightarrow\alpha}
\delta\mathcal{D}^-_{\beta\rightarrow\alpha}\\
&\approx
e^{i(2\textbf{k}_L\cdot(\textbf{R}_{\beta}-\textbf{R}_{\alpha})-2kR)}
\chi_{\sst ++}\chi_{\sst
-+}\mathcal{D}^-_{\beta}\mathcal{D}^-_{\beta}
+\chi_{\sst ++}\chi_{\sst --}\mathcal{D}^-_{\beta}\mathcal{D}^+_{\beta}\\
&\phantom{\approx} +\chi_{\sst +-}\chi_{\sst
-+}\mathcal{D}^+_{\beta}\mathcal{D}^-_{\beta}
+e^{i(2\textbf{k}_L\cdot(\textbf{R}_{\alpha}-\textbf{R}_{\beta})+2kR)}
\chi_{\sst +-}\chi_{\sst
--}\mathcal{D}^+_{\beta}\mathcal{D}^+_{\beta}.
\end{aligned}
\end{equation}
Averaging these expressions over the positions $\textbf{R}_\alpha$
and $\textbf{R}_\beta$ of the atoms while keeping $R\gg \lambda$
fixed, only terms with position-independent phases
survive, giving rise to:
\begin{equation}
\begin{aligned}
C^{(2)}&\approx\chi_{\sst +-}\chi_{\sst
-+}\,\mathcal{D}^+_{\alpha}
\mathcal{D}^-_{\beta}\\
L^{(2)}&\approx\chi_{\sst ++}\chi_{\sst --}\,\mathcal{D}^-_{\beta}
\mathcal{D}^+_{\beta}+ \chi_{\sst +-}\chi_{\sst
-+}\,\mathcal{D}^+_{\beta}\mathcal{D}^-_{\beta}.
\end{aligned}
\end{equation}

This simple model allows to understand clearly why there are twice
more terms in the ladder expression than in the crossed one.
Fields generated in the FFWM process \emph{always} interfere
constructively in the case of the ladder, since they originate
from the same atom. Of course, in the preceding explanation, we
have discarded polarization effects and inelastic processes in the
nonlinear susceptibilities. Nevertheless, even if in that case the
situation becomes more involved, the differences between the
ladder and crossed expressions still arise from this local four
wave-mixing process. For example, in the last line of
Eqs.~\eqref{corrcbscross} and \eqref{corrcbsladder}, we see that
the operator
$\big(G_{ij}^{\alpha^+_q}[\Delta]X_j^{\alpha^{(0)}}\otimes\big)$
plays the role of a generalized nonlinear susceptibility
(actually, the standard ones are recovered from the elastic part
of $X_j^{\alpha^{(0)}}$). Thus we recover the same structure as
previously depicted, which leads to similar conclusions.

Finally, as mentioned above, for large saturation parameters
$s_0$, even if in that case the total scattered intensities
(ladder and crossed) are dominated by the inelastic spectrum, we
numerically observe that the enhancement factor does not vanish
but rather goes to a finite limit $1.096$ (for $\delta=0$). Field
coherence is thus 
not fully erased, which, at first glance, could be surprising
since the inelastic spectrum is a noise spectrum at the heart of
the temporal decoherence of the radiated field. But this only
means that both crossed and ladder become vanishingly small
relatively to the incident intensity. Nevertheless, even if it
would be hard to derive it analytically from
Eqs.~\eqref{corrcbscross} and \eqref{corrcbsladder}, they actually
decrease at the same rate, resulting in a finite (but small)
enhancement factor.

\section{Conclusion}

In the case of two atoms, even if the quantum Langevin approach
leads to calculations more tedious and involved than the direct
optical Bloch method, it nevertheless gives rise to an
understanding closer to the usual scattering approach developed in
the linear regime. In this way, one also gets direct information
about the inelastic spectrum of the radiated light. In particular,
it clearly outlines the crucial roles played by the inelastic
nonlinear susceptibilities and by the quantum correlations of the
vacuum fluctuations. Furthermore, since the framework of the
quantum Langevin approach is set in the frequency domain,
frequency-dependent propagation (\emph{i.e.} frequency-dependent
mean-free paths) between the atoms can be naturally included.

The next step would be to adapt the present approach to
"macroscopic" configurations (\emph{i.e.} at least many atoms),
allowing for a more direct comparison with existing
experiments~\cite{thierry}. This would provide a better
understanding of light transport properties in nonlinear atomic
media where vacuum fluctuations play a role. In particular, for
given values of the incident laser intensity and detuning, the
nonlinear mean-free path becomes negative in well-defined
frequency windows. This means that light \textit{amplification}
can be achieved in these frequency windows~\cite{pra5M,prl38WEDM}.
The atomic media would then constitute a very simple realization
of a coherent random laser.

\begin{acknowledgments}
We would like to thank Cord~M\"uller, Oliver~Sigwarth, Andreas
Buchleitner, Vyacheslav Shatokhin, Serge Reynaud and Jean-Michel
Courty for stimulating discussions. T.W. has been supported by the
DFG Emmy Noether program. Laboratoire Kastler Brossel is
laboratoire de l'Universit\'e Pierre et Marie Curie et de l'Ecole
Normale Sup\'erieure, UMR 8552 du CNRS.
\end{acknowledgments}

\appendix

\section{}
\label{ggbar}
The $g\bar{g}$ terms in Eq.~\eqref{corrggb} read:
\begin{multline}
{X}_{i'}^{\beta}[\Delta']{X}_i^{\alpha}[\Delta]=\cdots\\
-g\bar{g}\biggl\{ {X}_{i'}^{\beta}[\Delta']
\biggl[G_{ij}^{\alpha^+_q}[\Delta]\left(X_j^{\alpha^{(0)}} \otimes
G^{\beta^-_p}_{\mathcal{D}_q^-j'}
\left(\mathcal{D}^{{\alpha+}^{(0)}}_p \otimes
X^{\beta^{(0)}}_{j'}\right)\right)[\Delta]
+G_{ij}^{\alpha^+_q}[\Delta]\left(G^{\alpha^-_p}_{jj'}
\left(\mathcal{D}^{{\beta+}^{(0)}}_p\otimes
X_{j'}^{\alpha^{(0)}}\right)\otimes
\mathcal{D}^{{\beta-}^{(0)}}_q\right)[\Delta]
\biggr.\biggr.\\
\biggl.+G_{ij}^{\alpha^-_q}[\Delta]
\left(\mathcal{D}^{{\beta+}^{(0)}}_q\otimes
G^{\alpha^+_p}_{jj'}\left(X_{j'}^{\alpha^{(0)}}\otimes
\mathcal{D}^{{\beta-}^{(0)}}_p\right)\right)[\Delta]
+G_{ij}^{\alpha^-_q}[\Delta]\left(G^{\beta^+_p}_{\mathcal{D}_q^-j'}
\left(X_{j'}^{\beta^{(0)}}\otimes
\mathcal{D}^{{\alpha-}^{(0)}}_p\right)\otimes
X_j^{\alpha^{(0)}}\right)[\Delta]
\biggr]\\
\biggl[ G_{i'j'}^{\beta^+_q}[\Delta']\left(X_{j'}^{\beta^{(0)}}
\otimes G^{\alpha^-_p}_{\mathcal{D}_q^-j}
\left(\mathcal{D}^{{\beta+}^{(0)}}_p \otimes
X^{\alpha^{(0)}}_{j}\right)\right)[\Delta']
+G_{i'j'}^{\beta^+_q}[\Delta']\left(G^{\beta^-_p}_{j'j}
\left(\mathcal{D}^{{\alpha+}^{(0)}}_p\otimes
X_{j}^{\beta^{(0)}}\right)\otimes
\mathcal{D}^{{\alpha-}^{(0)}}_q\right)[\Delta']
\biggr.\\
\biggl.+G_{i'j'}^{\beta^-_q}[\Delta']
\left(\mathcal{D}^{{\alpha+}^{(0)}}_q\otimes
G^{\beta^+_p}_{j'j}\left(X_{j}^{\beta^{(0)}}\otimes
\mathcal{D}^{{\alpha-}^{(0)}}_p\right)\right)[\Delta']
+G_{i'j'}^{\beta^-_q}[\Delta']\left(G^{\alpha^+_p}_{\mathcal{D}_q^-j}
\left(X_{j}^{\alpha^{(0)}}\otimes
\mathcal{D}^{{\beta-}^{(0)}}_p\right)\otimes
X_{j'}^{\beta^{(0)}}\right)[\Delta'] \biggr]
{X}_i^{\alpha^{(0)}}[\Delta]\\
+\biggl[G_{i'j'}^{\beta^+_p}[\Delta']
\biggl(X_{j'}^{\beta^{(0)}}\otimes
\mathcal{D}^{{\alpha-}^{(0)}}_p\biggr)[\Delta']\biggr]
\biggl[G_{ij}^{\alpha^-_q}[\Delta]
\biggl(\mathcal{D}^{{\beta+}^{(0)}}_q\otimes
X_j^{\alpha^{(0)}}\biggr)[\Delta]\biggr]\\
+\biggl.\biggl[G_{i'j'}^{\beta^-_p}[\Delta']
\biggl(\mathcal{D}^{{\alpha+}^{(0)}}_p\otimes
X_{j'}^{\beta^{(0)}}\biggr)[\Delta']\biggr]
\biggl[G_{ij}^{\alpha^+_q}[\Delta]
\biggl(X_j^{\alpha^{(0)}}\otimes
\mathcal{D}^{{\beta-}^{(0)}}_q\biggr)[\Delta] \biggr]\biggr\}
\end{multline}
\begin{multline}
{X}_{i'}^{\alpha}[\Delta']{X}_i^{\alpha}[\Delta]=\cdots\\
-g\bar{g}\biggl\{ {X}_{i'}^{\alpha}[\Delta']
\biggl[G_{ij}^{\alpha^+_q}[\Delta]\left(X_j^{\alpha^{(0)}} \otimes
G^{\beta^-_p}_{\mathcal{D}_q^-j'}
\left(\mathcal{D}^{{\alpha+}^{(0)}}_p \otimes
X^{\beta^{(0)}}_{j'}\right)\right)[\Delta]
+G_{ij}^{\alpha^+_q}[\Delta]\left(G^{\alpha^-_p}_{jj'}
\left(\mathcal{D}^{{\beta+}^{(0)}}_p\otimes
X_{j'}^{\alpha^{(0)}}\right)\otimes
\mathcal{D}^{{\beta-}^{(0)}}_q\right)[\Delta]
\biggr.\biggr.\\
\biggl.+G_{ij}^{\alpha^-_q}[\Delta]
\left(\mathcal{D}^{{\beta+}^{(0)}}_q\otimes
G^{\alpha^+_p}_{jj'}\left(X_{j'}^{\alpha^{(0)}}\otimes
\mathcal{D}^{{\beta-}^{(0)}}_p\right)\right)[\Delta]
+G_{ij}^{\alpha^-_q}[\Delta]\left(G^{\beta^+_p}_{\mathcal{D}_q^-j'}
\left(X_{j'}^{\beta^{(0)}}\otimes
\mathcal{D}^{{\alpha-}^{(0)}}_p\right)\otimes
X_j^{\alpha^{(0)}}\right)[\Delta]\biggr]\\
\biggl[ G_{i'j'}^{\alpha^+_q}[\Delta']\left(X_{j'}^{\alpha^{(0)}}
\otimes G^{\beta^-_p}_{\mathcal{D}_q^-j}
\left(\mathcal{D}^{{\alpha+}^{(0)}}_p \otimes
X^{\beta^{(0)}}_{j}\right)\right)[\Delta']
+G_{i'j'}^{\alpha^+_q}[\Delta']\left(G^{\alpha^-_p}_{j'j}
\left(\mathcal{D}^{{\beta+}^{(0)}}_p\otimes
X_{j}^{\alpha^{(0)}}\right)\otimes
\mathcal{D}^{{\beta-}^{(0)}}_q\right)[\Delta']\biggr.\\
\biggl.+G_{i'j'}^{\alpha^-_q}[\Delta']
\left(\mathcal{D}^{{\beta+}^{(0)}}_q\otimes
G^{\alpha^+_p}_{j'j}\left(X_{j}^{\alpha^{(0)}}\otimes
\mathcal{D}^{{\beta-}^{(0)}}_p\right)\right)[\Delta']
+G_{i'j'}^{\alpha^-_q}[\Delta']\left(G^{\beta^+_p}_{\mathcal{D}_q^-j}
\left(X_{j}^{\beta^{(0)}}\otimes
\mathcal{D}^{{\alpha-}^{(0)}}_p\right)\otimes
X_{j'}^{\alpha^{(0)}}\right)[\Delta']\biggr]{X}_i^{\alpha^{(0)}}[\Delta]\\
+\biggl[G_{i'j'}^{\alpha^+_p}[\Delta']
\biggl(X_{j'}^{\alpha^{(0)}}\otimes
\mathcal{D}^{{\beta-}^{(0)}}_p\biggr)[\Delta']\biggr]
\biggl[G_{ij}^{\alpha^-_q}[\Delta]
\biggl(\mathcal{D}^{{\beta+}^{(0)}}_q\otimes
X_j^{\alpha^{(0)}}\biggr)[\Delta]\biggr]\\
+\biggl.\biggl[G_{i'j'}^{\alpha^-_p}[\Delta']
\biggl(\mathcal{D}^{{\beta+}^{(0)}}_p\otimes
X_{j'}^{{\alpha}^{(0)}}\biggr)[\Delta']\biggr]
\biggl[G_{ij}^{\alpha^+_q}[\Delta]
\biggl(X_j^{\alpha^{(0)}}\otimes
\mathcal{D}^{{\beta-}^{(0)}}_q\biggr)[\Delta] \biggr]\biggr\}
\end{multline}

\section{Three-body correlation functions}
\label{matdthreebody}

\subsection{Single atom case}

The three-body correlation function for the Langevin force reads:
\begin{equation}
C_{abc}[\Delta',\Delta]=
\frac{1}{2\pi}\iint d\Delta_1d\Delta_2\delta[\Delta_1+\Delta_2-\Delta]
f[\Delta_1]g[\Delta_2]
\left\langle F^{{\alpha}}_{a}[\Delta']F^{{\alpha}}_{b}[\Delta_1]
F^{{\alpha}}_{c}[\Delta_2]\right\rangle,
\end{equation}
where $f[\Delta]$ and $g[\Delta]$ are regular functions such that
the preceding integral is well defined. Going back to the time
domain, $C_{abc}[\Delta',\Delta]$ reads as follows:
\begin{equation}
C_{abc}[\Delta',\Delta]=\frac{1}{2\pi}\iint dt dt' e^{i\Delta
t}e^{i\Delta't'} \iiiint dt_1dt_2dt_3dt_4
\delta(t_1+t_2-t)\delta(t_3+t_4-t)f(t_1)g(t_3) \left\langle
F^{{\alpha}}_{a}(t')F^{{\alpha}}_{b}(t_2)
F^{{\alpha}}_{c}(t_4)\right\rangle.
\end{equation}
Then, from the time correlation properties of the vacuum field,
one can show that:
\begin{equation}
\begin{aligned}
\left\langle F^{{\alpha}}_{a}(t')F^{{\alpha}}_{b}(t_2)
F^{{\alpha}}_{c}(t_4)\right\rangle&=4T^{q+}_{aa'}T^{q-}_{bb'}\delta(t'-t_2)
\left\langle X^{{\alpha}}_{a'}(t')X^{{\alpha}}_{b'}(t')
F^{{\alpha}}_{c}(t_4)\right\rangle\\
&+4T^{q+}_{aa'}T^{q-}_{cc'}\delta(t'-t_4)
\left\langle X^{{\alpha}}_{a'}(t')F^{{\alpha}}_{b}(t_2)
X^{{\alpha}}_{c'}(t_4)\right\rangle\\
&+4T^{q+}_{bb'}T^{q-}_{cc'}\delta(t_2-t_4)
\left\langle F^{{\alpha}}_{a}(t')X^{{\alpha}}_{b'}(t_2)
X^{{\alpha}}_{c'}(t_2)\right\rangle.
\end{aligned}
\end{equation}
where the $T^{q\pm}$ are $15\times15$ matrices defined by
$\left[X_i,\mathcal{D}^{\pm}_q\right]=\pm 2T^{q\pm}_{ij}X_j$.

When taken at the same time, the atomic operators (including the
identity $\openone$) define a group entirely characterized
by the group structure constants $\epsilon_{ij}^{\phantom{ij}k}$,
\emph{i.e.}:
\begin{equation}
X_i(t)X_j(t)=\sum_k\epsilon_{ij}^{\phantom{ij}k}X_k(t),
\end{equation}
so that the preceding equation becomes:
\begin{equation}
\begin{aligned}
\left\langle F^{{\alpha}}_{a}(t')F^{{\alpha}}_{b}(t_2)
F^{{\alpha}}_{c}(t_4)\right\rangle&=4T^{q+}_{aa'}T^{q-}_{bb'}\delta(t'-t_2)
\epsilon_{a'b'}^{\phantom{a'b'}u}\left\langle X^{{\alpha}}_{u}(t')
F^{{\alpha}}_{c}(t_4)\right\rangle\\
&+4T^{q+}_{aa'}T^{q-}_{cc'}\delta(t'-t_4)
\left\langle X^{{\alpha}}_{a'}(t')F^{{\alpha}}_{b}(t_2)
X^{{\alpha}}_{c'}(t_4)\right\rangle\\
&+4T^{q+}_{bb'}T^{q-}_{cc'}\delta(t_2-t_4)
\epsilon_{a'b'}^{\phantom{b'c'}u}
\left\langle F^{{\alpha}}_{a}(t')X^{{\alpha}}_{u}(t_2)\right\rangle.
\end{aligned}
\end{equation}
Injecting the preceding relations in $C(a,b,c)$ and going back to
the frequency domain, we get:
\begin{equation}
\begin{aligned}
C_{abc}[\Delta',\Delta]&=4T^{q+}_{aa'}T^{q-}_{bb'}\epsilon_{a'b'}^{\phantom{a'b'}u}
\frac{1}{2\pi}\iint d\Delta_1d\Delta_2\delta(\Delta_1+\Delta_2-\Delta)
f[\Delta_1]g[\Delta_2]\left\langle X^{{\alpha}}_{u}[\Delta'+\Delta_1]
F^{{\alpha}}_{c}[\Delta_2]\right\rangle\\
&\phantom{=}+4T^{q+}_{aa'}T^{q-}_{cc'}\frac{1}{2\pi}\int d\Delta_3
g[\Delta_3]f[\Delta-\Delta_3]
D^{b,\alpha\alpha\alpha}_{a'c'}[\Delta'+\Delta_3,\Delta-\Delta_3]\\
&\phantom{=}+4T^{q+}_{bb'}T^{q-}_{cc'}\epsilon_{a'b'}^{\phantom{b'c'}u}
\left\langle F^{{\alpha}}_{a}[\Delta']X^{{\alpha}}_{u}[\Delta]\right\rangle
\frac{1}{2\pi}\iint d\Delta_1d\Delta_2\delta(\Delta_1+\Delta_2-\Delta)
f[\Delta_1]g[\Delta_2]\\
&=4T^{q+}_{aa'}T^{q-}_{bb'}\epsilon_{a'b'}^{\phantom{a'b'}u}
\frac{1}{2\pi}\iint d\Delta_1d\Delta_2\delta(\Delta_1+\Delta_2-\Delta)
f[\Delta_1]g[\Delta_2]G^{\alpha}_{uv}[\Delta'+\Delta_1]
\left\langle F^{{\alpha}}_{v}[\Delta'+\Delta_1]
F^{{\alpha}}_{c}[\Delta_2][\right\rangle\\
&\phantom{=}+4T^{q+}_{aa'}T^{q-}_{cc'}\frac{1}{2\pi}\int d\Delta_3
g[\Delta_3]f[\Delta-\Delta_3]
D^{b,\alpha\alpha\alpha}_{a'c'}[\Delta'+\Delta_3,\Delta-\Delta_3]\\
&\phantom{=}+4T^{q+}_{bb'}T^{q-}_{cc'}\epsilon_{a'b'}^{\phantom{b'c'}u}
G_{uv}^{\alpha}[\Delta]
\left\langle F^{{\alpha}}_{a}[\Delta']F^{{\alpha}}_{v}[\Delta]\right\rangle
\frac{1}{2\pi}\iint d\Delta_1d\Delta_2\delta(\Delta_1+\Delta_2-\Delta)
f[\Delta_1]g[\Delta_2]\\
&=2\pi\delta[\Delta+\Delta']4T^{q+}_{aa'}T^{q-}_{bb'}
\epsilon_{a'b'}^{\phantom{a'b'}u}D^{\alpha\alpha}_{vc}
\frac{1}{2\pi}\iint d\Delta_1d\Delta_2\delta(\Delta_1+\Delta_2-\Delta)
f[\Delta_1]g[\Delta_2]G^{\alpha}_{uv}[-\Delta_2]\\
&\phantom{=}+4T^{q+}_{aa'}T^{q-}_{cc'}\frac{1}{2\pi}\int d\Delta_3
g[\Delta_3]f[\Delta-\Delta_3]
D^{b,\alpha\alpha\alpha}_{a'c'}[\Delta'+\Delta_3,\Delta-\Delta_3]\\
&\phantom{=}+2\pi\delta[\Delta+\Delta']4T^{q+}_{bb'}T^{q-}_{cc'}
\epsilon_{a'b'}^{\phantom{b'c'}u}D^{\alpha\alpha}_{av}
G_{uv}^{\alpha}[\Delta]
\frac{1}{2\pi}\iint d\Delta_1d\Delta_2\delta(\Delta_1+\Delta_2-\Delta)
f[\Delta_1]g[\Delta_2],
\end{aligned}
\end{equation}
where we have introduced the matrix
$D^{b,\alpha\alpha\alpha}_{ik}[\Delta',\Delta]$ defined by:
\begin{equation}
D^{b,\alpha\alpha\alpha}_{ik}[\Delta',\Delta]=\frac{1}{2\pi}
\iint d\Delta_1d\Delta_2\delta(\Delta_1+\Delta_2-\Delta')
\left\langle X^{{\alpha}}_{i}[\Delta_1]F^{{\alpha}}_{b}[\Delta]
X^{{\alpha}}_{k}[\Delta_2]\right\rangle.
\end{equation}

This matrix is calculated using the same strategy (\emph{i.e.}
going back and forth to the time domain) and one finally gets:
\begin{multline}
D^{b,\alpha\alpha\alpha}_{ik}[\Delta',\Delta]=2\pi\delta[\Delta+\Delta']
\left\{G^{\alpha}_{ia}[0]L^{\alpha}_a
G^{\alpha}_{kc}[\Delta']\tilde{D}^{\alpha\alpha}_{bc}+
G^{\alpha}_{ia}[\Delta']G^{\alpha}_{kc}[0]
L^{\alpha}_c\tilde{D}^{\alpha\alpha}_{ab}\right.\\
\left.+4T^{q+}_{bb'}T^{q-}_{cc'}\epsilon_{b'c'}^{\phantom{b'c'}v}
\tilde{D}^{\alpha\alpha}_{au}
\frac{1}{2\pi}\iint d\Delta_1d\Delta_2\delta(\Delta_1+\Delta_2-\Delta')
G^{\alpha}_{ia}[\Delta_1]G^{\alpha}_{kc}[\Delta_2]
G^{\alpha}_{vu}[-\Delta_1]\right.\\
\left.+4T^+_{aa'}T^-_{bb'}\epsilon_{a'b'}^{\phantom{a'b'}v}
\tilde{D}^{\alpha\alpha}_{uc}
\frac{1}{2\pi}\iint d\Delta_1d\Delta_2\delta(\Delta_1+\Delta_2-\Delta')
G^{\alpha}_{ia}[\Delta_1]G^{\alpha}_{kc}[\Delta_2]
G^{\alpha}_{vu}[-\Delta_2]
\right\}\\
+4T^{q+}_{aa'}T^{q-}_{cc'}
\left(\frac{1}{2\pi}\iint
d\Delta_3d\Delta_4\delta(\Delta_3+\Delta_4-\Delta')
G^{\alpha}_{ia}[\Delta_3]G^{\alpha}_{kc}[\Delta_4]\right)\\
\times\left(\frac{1}{2\pi}\iint
d\Delta_1d\Delta_2\delta(\Delta_1+\Delta_2-\Delta')
\left\langle X^{{\alpha}}_{a'}[\Delta_1]F^{{\alpha}}_{b}[\Delta]
X^{{\alpha}}_{c'}[\Delta_2]\right\rangle\right).
\end{multline}

It may seem that we have taken a loop path and that we are back to
square one... However, in the last line of the preceding formula,
we immediately recognize the matrix
$D^{b,\alpha\alpha\alpha}_{a'b'}[\Delta',\Delta]$. Thus, the
preceding equation is nothing else but a linear system for this
matrix. More precisely,
$D^{b,\alpha\alpha\alpha}_{ik}[\Delta',\Delta]$ is the solution of
the following linear system:
\begin{equation}
\label{linear}
D_{ik}^{b,\alpha\alpha\alpha}[\Delta',\Delta]
-I_{ik,a'c'}^{\alpha\alpha}[\Delta']
D_{a'c'}^{b,\alpha\alpha\alpha}[\Delta',\Delta]=
J_{ik}^{b,\alpha\alpha\alpha}[\Delta',\Delta],
\end{equation}
with
\begin{equation}
\left\{\begin{aligned}
I_{ik,a'c'}^{\alpha\alpha}[\Delta']&=4T^{q+}_{aa'}T^{q-}_{cc'}
\frac{1}{2\pi}\iint d\Delta_3d\Delta_4\delta(\Delta_3+\Delta_4-\Delta')
G^{\alpha}_{ia}[\Delta_3]G^{\alpha}_{kc}[\Delta_4]\\
J_{ik}^{b,\alpha\alpha\alpha}[\Delta',\Delta]&=
2\pi\delta[\Delta+\Delta']
\left\{G^{\alpha}_{ia}[0]L^{\alpha}_a
G^{\alpha}_{kc}[\Delta']\tilde{D}^{\alpha\alpha}_{bc}+
G^{\alpha}_{ia}[\Delta']G^{\alpha}_{kc}[0]
L^{\alpha}_c\tilde{D}^{\alpha\alpha}_{ab}\right.\\
&\phantom{=2\pi\delta[\Delta+\Delta']}\quad
\left.+4T^{q+}_{bb'}T^{q-}_{cc'}\epsilon_{b'c'}^{\phantom{b'c'}v}
\tilde{D}^{\alpha\alpha}_{au}
\frac{1}{2\pi}\iint d\Delta_1d\Delta_2\delta(\Delta_1+\Delta_2-\Delta')
G^{\alpha}_{ia}[\Delta_1]G^{\alpha}_{kc}[\Delta_2]
G^{\alpha}_{vu}[-\Delta_1]\right.\\
&\phantom{=2\pi\delta[\Delta+\Delta']}\quad
\left.+4T^+_{aa'}T^-_{bb'}\epsilon_{a'b'}^{\phantom{a'b'}v}
\tilde{D}^{\alpha\alpha}_{uc}
\frac{1}{2\pi}\iint d\Delta_1d\Delta_2\delta(\Delta_1+\Delta_2-\Delta')
G^{\alpha}_{ia}[\Delta_1]G^{\alpha}_{kc}[\Delta_2]
G^{\alpha}_{vu}[-\Delta_2]
\right\}
\end{aligned}\right..
\end{equation}

In the preceding equations, the Green's function $G[\Delta]$ and
the diffusion matrix $D^{\alpha\alpha}$ only depend on the Rabi
field $\Omega_L$ evaluated at the position of atom $\alpha$. Thus,
for \emph{any} value of $\Delta$, numerical values of $I$ and $J$
can be computed, allowing for a direct calculation of
$D_{ik}^{b,\alpha\alpha\alpha}[-\Delta,\Delta]$. Furthermore, it
is not surprising that the matrix $I$ shows up in the linear
system. Indeed, the Green's function $G[\Delta]$ governs the time
evolution of $\textbf{X}$ through a Fourier transform. Thus the
time evolution of products of operators
$\textbf{X}_i(t)\textbf{X}_j(t)$ will be simply governed by the
Fourier transform of the product of two Green's functions
$G(t)G(t)$, which is precisely the convolution product found in
$I$. Finally, from the knowledge of the matrix $D$, we can
calculate the value of $C_{abc}[\Delta',\Delta]$:
\begin{equation}
\begin{aligned}
C_{abc}[\Delta',\Delta]&=2\pi\delta[\Delta+\Delta']
\left\{4T^{q+}_{aa'}T^{q-}_{bb'}
\epsilon_{a'b'}^{\phantom{a'b'}u}D^{\alpha\alpha}_{vc}
\frac{1}{2\pi}\iint d\Delta_1d\Delta_2\delta(\Delta_1+\Delta_2-\Delta)
f[\Delta_1]g[\Delta_2]G^{\alpha}_{uv}[-\Delta_2]\right.\\
&\phantom{=2\pi\delta[\Delta+\Delta']}\quad
+4T^{q+}_{aa'}T^{q-}_{cc'}
\frac{1}{2\pi}\iint d\Delta_1d\Delta_2\delta(\Delta_1+\Delta_2-\Delta)
f[\Delta_1]g[\Delta_2]
D^{b,\alpha\alpha\alpha}_{a'c'}[-\Delta_1,\Delta_1]\\
&\phantom{=2\pi\delta[\Delta+\Delta']}\quad
+\left.4T^{q+}_{bb'}T^{q-}_{cc'}
\epsilon_{a'b'}^{\phantom{b'c'}u}D^{\alpha\alpha}_{av}
G_{uv}^{\alpha}[\Delta]
\frac{1}{2\pi}\iint d\Delta_1d\Delta_2\delta(\Delta_1+\Delta_2-\Delta)
f[\Delta_1]g[\Delta_2]\right\}.
\end{aligned}
\end{equation}
Of course, we recover the global factor
$2\pi\delta[\Delta+\Delta']$, showing that the time correlation
function only depends on the time difference $t'-t$ (stationary
condition).

\subsection{Two-atom case}

The calculation of quantities like:
\begin{equation}
C^{\alpha\beta}_{abc}[\Delta',\Delta]= \frac{1}{2\pi}\iint
d\Delta_1d\Delta_2\delta[\Delta_1+\Delta_2-\Delta]
f[\Delta_1]g[\Delta_2] \left\langle
F^{\alpha}_{j'}[\Delta']F^{\beta}_{k}[\Delta_1]
F^{{\alpha}}_{k'}[\Delta_2]\right\rangle^{(\bar{g})},
\end{equation}
follows, more or less, the way described in the preceding section.
In particular, it also involves the calculation of a matrix
$D^{b,{\alpha}\beta\alpha^{(\bar{g})}}_{ik}[\Delta',\Delta]$
defined as follows:
\begin{equation}
D^{b,{\alpha}\beta\alpha^{(\bar{g})}}_{ik}[\Delta',\Delta]=
\frac{1}{2\pi} \iint
d\Delta_1d\Delta_2\delta(\Delta_1+\Delta_2-\Delta') \left\langle
X^{{\alpha}}_{i}[\Delta_1]F^{\beta}_{b}[\Delta]
X^{{\alpha}}_{k}[\Delta_2]\right\rangle^{(\bar{g})}.
\end{equation}
The latter is also found to be the solution of a linear system,
resembling the preceding one (see Eq.~\eqref{linear}):
\begin{equation}
D_{ik}^{b,{\alpha}\beta\alpha^{(\bar{g})}}[\Delta',\Delta]
-I_{ik,a'c'}^{{\alpha}{\alpha}}[\Delta']
D_{a'c'}^{b,{\alpha}\beta\alpha^{(\bar{g})}}[\Delta',\Delta]=
J_{ik}^{b,{\alpha}\beta\alpha^{(\bar{g})}}[\Delta',\Delta],
\end{equation}
with
\begin{multline}
J_{ik}^{b,\alpha\beta\alpha^{(\bar{g})}}[\Delta',\Delta]=
-\left(\frac{1}{2}\right)2\pi\delta[\Delta+\Delta']\biggl\{
G^{\alpha}_{ia}[0]L^{\alpha}_a
G^{\alpha}_{kc}[\Delta']\tilde{D}^{\beta\alpha^{(0)}}_{bc}
+G^{\alpha}_{ia}[\Delta']G^{\alpha}_{kc}[0]
L^{\alpha}_c\tilde{D}^{\alpha\beta^{(0)}}_{ab}\biggr.\\
+4T^{q+}_{bb'}\langle X^{\beta^{(0)}}_{b'}\rangle
\frac{1}{2\pi}\iint
d\Delta_1d\Delta_2\delta(\Delta_1+\Delta_2-\Delta')
G^{\alpha}_{ia}[\Delta_1]G^{\alpha^-_q}_{kc}[\Delta_2]
G^{\alpha}_{cu}[-\Delta_1]\tilde{D}^{\alpha\alpha^{(0)}}_{au}\\
+4T^{q-}_{bb'}\langle X^{\beta^{(0)}}_{b'}\rangle
\frac{1}{2\pi}\iint
d\Delta_1d\Delta_2\delta(\Delta_1+\Delta_2-\Delta')
G^{\alpha^+_q}_{ia}[\Delta_1]G^{\alpha}_{kc}[\Delta_2]
G^{\alpha}_{au}[-\Delta_2]\tilde{D}^{\alpha\alpha^{(0)}}_{uc}\\
-2G^{\beta}_{\mathcal{D}^+_qu}[\Delta']
\tilde{D}^{\beta\beta^{(0)}}_{ub} \frac{1}{2\pi}\iint
d\Delta_1d\Delta_2\delta(\Delta_1+\Delta_2-\Delta')
G^{\alpha^-_q}_{ia}[\Delta_1]\langle\tilde{X}^{\alpha^{(0)}}_a[-\Delta_2]
\tilde{X}^{\alpha^{(0)}}_k[\Delta_2]\rangle^{(0)}\\
\biggl. -2G^{\beta}_{\mathcal{D}^+_qu}[\Delta']
\tilde{D}^{\beta\beta^{(0)}}_{bu} \frac{1}{2\pi}\iint
d\Delta_1d\Delta_2\delta(\Delta_1+\Delta_2-\Delta')
G^{\alpha^-_q}_{kc}[\Delta_2]\langle\tilde{X}^{\alpha^{(0)}}_i[\Delta_1]
\tilde{X}^{\alpha^{(0)}}_c[-\Delta_1]\rangle^{(0)}\biggr\}.
\end{multline}


\begin{thebibliography}{999}
\bibitem{coherence} Please note however that this result is no longer true as soon as inelastic scattering occurs in a medium : in this case, CBS
  can arise from a \emph{three-wave} interference \cite{facteur3}.
\end{thebibliography}
\end{document}